\documentclass[journal,10pt]{IEEEtran}

\usepackage{caption}
\usepackage{subcaption}
\usepackage{amssymb}
\usepackage{hyperref}
\usepackage{amsmath}
\usepackage{graphicx}
\usepackage{multirow}
\usepackage{tabularx}
\usepackage{float, caption, booktabs, cellspace}
\usepackage{bm}
\usepackage{color}
\usepackage[shortlabels]{enumitem}
\usepackage{lipsum}

\setlist[enumerate]{wide=\parindent}
\usepackage{rotating}

\usepackage[table]{xcolor}

\usepackage{menukeys}
\makeatletter
\DeclareRobustCommand{\rvdots}{%
  \vbox{
    \baselineskip3\p@\lineskiplimit\z@
    \kern2.5\p@
    \hbox{.}\hbox{.}\hbox{.}
  }}
\makeatother

\begin{document}

\title{A Supervised Learning Framework for Joint Angle-of-Arrival and Source Number Estimation}

\author{Noud~Kanters~and~Andrés~Alayón~Glazunov,~\IEEEmembership{Senior~Member,~IEEE}

\thanks{Noud Kanters is with the Department of Electrical Engineering, University of Twente, 7500 AE Enschede, The Netherlands (e-mail: n.b.kanters@utwente.nl).}% <-this % stops a space
\thanks{Andrés Alayón Glazunov is with the Department of Electrical Engineering, University of Twente, 7500 AE Enschede, The Netherlands, and also with the Department of Electrical Engineering, Chalmers University of Technology, 412 96 Gothenburg, Sweden (e-mail: a.alayonglazunov@utwente.nl; andres.glazunov@chalmers.se).}}

% \markboth{IEEE Transactions on Signal Processing,~Vol.~XX, No.~XX, MMM~YYYY}
{}

% \author{Noud Kanters\IEEEauthorrefmark{1}, 
%         Andrés Alayón Glazunov\IEEEauthorrefmark{1}\IEEEauthorrefmark{2}\\
% 		\IEEEauthorrefmark{1}University of Twente, Department of Electrical Engineering, Enschede, Netherlands.\\
% 		\IEEEauthorrefmark{2}Chalmers University of Technology, Department of Electrical Engineering, Gothenburg, Sweden.\\
% 	\textit{Email: \{n.b.kanters, a.alayonglazunov\}@utwente.nl} }
% <-this % stops a space
	
\maketitle

\begin{abstract}
    Machine learning is a promising technique for angle-of-arrival (AOA) estimation of waves impinging a sensor array. However, the majority of the methods proposed so far only consider a known, fixed number of impinging waves, i.e., a fixed source number. This paper proposes a machine-learning-based estimator designed for the case when the source number is variable and hence unknown a priori. The proposed estimator comprises a framework of single-label classifiers. Each classifier predicts if waves are present within certain randomly selected segments of the array's field of view (FOV), resulting from discretising the FOV with a certain (FOV) resolution. The classifiers' predictions are combined into a probabilistic angle spectrum, whereupon the source number and the AOAs are estimated jointly by applying a probability threshold whose optimal level is learned from data. The estimator's performance is assessed using a new performance metric: the joint AOA estimation success rate. Numerical simulations show that for low SNR (-10 dB), a low FOV resolution (2\textdegree) yields a higher success rate than a high resolution (1\textdegree), whereas the opposite applies for mid (0 dB) and high (10 dB) SNRs. In nearly all simulations, except one at low SNR and a high FOV resolution, the proposed estimator outperforms the MUSIC algorithm if the maximum allowed AOA estimation error is approximately equal to (or larger than) the FOV resolution.
    
    % It is observed that for $SNR=-10$ dB, a grid resolution of $2^\circ$ yields more accurate AOA estimates than a $1^\circ$ resolution, whereas the opposite applies to 0 and 10 dB SNRs. The proposed estimator yields more accurate estimates than the MUSIC algorithm if the MUSIC angle spectrum is evaluated with the same resolution as the resolution of the grid. 
    
\end{abstract}


\section{Introduction}
\label{sec:introduction}

    \IEEEPARstart{A}{ngle}-of-arrival (AOA) estimation of waves impinging a sensor array has been studied extensively as it has applications in various fields from array signal processing, e.g., wireless communications, radar and sonar \cite{krim1996two}. In many practical applications, the number of waves impinging the array (henceforth called the source number) is not constant, meaning it has to be estimated as well. Solutions to this problem can be categorised into separable and joint detection methods, indicating whether the source number is estimated prior to or simultaneously with the AOAs, respectively \cite{van2004optimum}.
    
    Conventional AOA estimators generally require a source number estimate prior to the AOA estimation, and hence they correspond to the separable detection category. The source number estimate can be obtained through model order estimators like, e.g., Akaike's information criterion (AIC) or the minimum description length (MDL) \cite{wax1985detection}. Beamformers, e.g., the Bartlett and the Capon beamformers, belong to the class of conventional estimators \cite{capon1969high}. Their resolution, i.e., their ability to resolve closely spaced sources, depends directly on the physical size of the array \cite{krim1996two}. This limitation does not apply to the subspace-based algorithms, e.g., multiple signal classification (MUSIC) \cite{schmidt1986multiple}, estimation of signal parameters via rotational invariance techniques (ESPRIT) \cite{roy1989esprit}, and variants thereof like root-MUSIC \cite{barabell1983improving}. However, these algorithms require the computationally expensive eigenvalue decomposition. Moreover, the resolution of the MUSIC algorithm deteriorates for highly correlated signals, whereas ESPRIT and root-MUSIC can only be applied in combination with particular array geometries \cite{krim1996two}. Maximum likelihood (ML) methods, e.g., \cite{stoica1989music, ziskind1988maximum}, do not suffer from these fundamental limitations. However, their computational complexity grows exponentially with the source number. In order to mitigate the aforementioned shortcomings, various sparsity-based approaches have been proposed, e.g., \cite{yang2012off, chen2018off, liu2012efficient}. While these methods can handle scenarios of unknown source numbers (i.e., joint detection), spurious sources are often present in the resulting power spectra \cite{yang2018sparse}.
    
    % Alternatively, joint detection --> sparsity based

    % \textcolor{red}{(Already in this paragraph the "problems" with traditional AoA estimators need to be listed clearly. Is it only accuracy? What other problems are there?)(I think that the introduction is rather lengthy. You shouldn't be describing the methods. You should be more concise. Here what you need is (i) to establish if there is a problem that you overcome with supervised learning and (ii) are there problems to be addressed in supervise problems.)}
    % The conventional estimators all rely in some way on an assumed data model. The accuracy of these type of estimators therefore heavily depends on the consistency between this model and reality \cite{krim1996two}. Furthermore, 
    Recently, supervised-learning-based AOA estimation algorithms have been proposed to further improve the accuracy and/or the computational efficiency. These algorithms learn a mapping between array outputs and AOAs from data directly. Hence, they do not require specific assumptions regarding the array geometry or the data model. The majority of these supervised-learning-based works are only applicable if the source number is fixed, henceforth referred to as scenario I. In other words, they can be considered part of the separable detection category, but they do not consider the source number detection itself. For example in \cite{khan2018angle}, the 2D AOA estimation (i.e., azimuth and elevation angle estimation) of a single source is performed by combining the conventional MUSIC algorithm with different learning algorithms, i.e., neural networks (NN), Gaussian processes (GP) and regression trees (RT). All of them consistently outperform the baseline MUSIC algorithm in terms of the average AOA estimation error, with improvements up to 50\% for GP and RT in particular high-SNR, low-elevation situations. Similarly, \cite{zhu2020two} considers the 2D AOA estimation of a single source through an ensemble of five convolutional neural networks (CNNs), \cite{kase2018doa} investigates the 1D AOA estimation of two sources through a deep neural network (DNN) and \cite{ahmed2020deep} proposes to emulate a large array through a DNN, whereupon the 1D single-source AOA is estimated using the MUSIC algorithm. In \cite{pastorino2005smart}, multi-source (2, 3 and 6 sources) 1D AOA estimation is performed by using a separate support vector machine (SVM) for the estimation of each AOA. Although this implies that the source number determines how many SVMs are required, again the source number detection itself is not considered.

        In many practical applications the number of sources is not constant (which we refer to as scenario II in this paper), hence it is to be estimated too. Therefore, the joint estimation of the source number and the AOAs, i.e., the alternative to separable detection, is of great relevance. Clearly, an AOA estimator performing joint estimation comes with increased complexity, as it should be capable of estimating a variable number of parameters. Since this is not straightforward to implement using existing learning algorithms, it has received less attention. Nevertheless, a number of solutions have been proposed. For example, in \cite{bialer2019performance}, a single DNN is deployed for the estimation of both the source number (restricted to be between 1 and 4 by design) and the 1D AOAs. More freedom in terms of the source numbers that can be handled is provided by the methods presented in \cite{liu2018direction} and \cite{papageorgiou2021deep}. There, the estimators (comprising multiple parallel DNNs in \cite{liu2018direction} and a single CNN in \cite{papageorgiou2021deep}) are tailored to a 1D grid of search angles (1\textdegree~resolution) within the FOV of the sensor array. Hence, they formulate the AOA estimation problem as a classification problem and aim to find those search angles which represent AOAs. In \cite{liu2018direction}, the predictions for all search angles are combined into an angle spectrum, whereupon the arguments of the highest peaks are returned as the AOA estimates. However, it is not explained how the estimator deals with scenarios of unknown source numbers. On the contrary, in \cite{papageorgiou2021deep} a user-defined confidence level is used to estimate the source number. However, this level is not optimized. Furthermore, neither \cite{liu2018direction} nor \cite{papageorgiou2021deep} investigates how the grid resolution itself affects the predictions of the used learning algorithms.  
        % Although solely trained on data representing 2 sources, it is shown that the estimator can handle scenarios with 1 and 3 sources as well.
        
        In this paper, we adopt an approach comparable to the ones presented in \cite{liu2018direction,papageorgiou2021deep}, i.e., we discretise the array's FOV, whereupon the joint AOA estimation problem is solved through classification. The main contributions of this paper can be summarized as follows:
        \begin{enumerate}[-,nosep]
            \item A machine learning framework (MLF) is proposed to jointly estimate the source number and the AOAs of waves impinging an sensor array. The MLF consists of an ensemble of classifiers, trained through supervised learning, which are organized along a framework based on the ensemble method random $k$-labelsets (RA$k$EL) \cite{tsoumakas2010random}. Consequently, the proposed MLF can, in principle, be deployed in combination with any learning algorithm capable of single-label multi-class classification. Modifications to the RA$k$EL method are implemented to tailor it to the AOA estimation problem.  
            % Results are specialized to a uniform (1D) FOV similar to the approach presented in \cite{liu2018direction}.   
            \item A peak detection algorithm is devised in order to jointly extract the source number and the AOAs from the probabilistic angle spectrum. This algorithm comprises a probability threshold, whose level is optimized based on data. The spectrum peaks above the threshold are located, whereupon the number of peaks and their arguments are returned as the source number estimate and the AOA estimates, respectively. 
        	\item The impact of the resolution of the FOV discretisation (FOV resolution) on the predictions of the individual classifiers as well as on the final AOA estimates is investigated through numerical simulations, using feedforward NNs as the learning algorithm. It is shown that increasing the FOV resolution does not necessarily improve the overall joint AOA estimation success rate (see next point) of the MLF, depending on the signal-to-noise ratio (SNR).
        	\item A new performance metric, the joint AOA estimation success rate, is introduced. This metric is based on the $success~rate$ proposed in \cite{zhu2020two}, but here we adapt it to take into account both the source number and the AOAs, and to make it depend on a user-defined maximum allowed AOA estimation error. Its theoretical upper bound (assuming ideal classifiers), imposed by the source number and the FOV resolution, is derived for the case of uniformly distributed random AOAs.
    	    \item The MLF is compared to the conventional MUSIC algorithm \cite{schmidt1986multiple} combined with the MDL and the AIC source number estimators \cite{wax1985detection}. Numerical simulations representing a variety of SNRs (-10, 0, 10 dB) and FOV resolutions (2\textdegree~and 1\textdegree) in both scenarios I and II show that the proposed MLF achieves a higher rate of successful joint AOA estimation than the MUSIC algorithm if the maximum allowed AOA estimation error is of the order of (or larger than) the FOV resolution. This applies to nearly all considered cases, except one at low SNR (-10 dB) and high FOV resolution (1\textdegree) in scenario I. 
        	
        % 	\textcolor{blue}{I introduce the text below, but I think it should be removed?}
        % 	\textcolor{red}{Assuming the same FOV resolution it is shown that the proposed MLF approach outperforms the MUSIC algorithm when the number of sources is variable, at both low ($-10$ dB) and high ($10$ dB) SNRs. This result depends on the maximum allowed AOA estimation error. At low SNR the MLF shows equal or better performance than the MUSIC independently on the maximum allowed AOA estimation error. However, as the SNR is increased the performance of the MLF is at least equal or better than the MUSIC algorithm above a certain maximum allowed AOA estimation error. For the scenario when the source number is fixed the performance behavior is similar. However, when higher FOV resolution is considered for lower SNR data, and just in this case, the performance of the MLF is somewhat worse than the MUSIC algorithm combined with the AIC. The MLF outperforms outperforms the MUSIC algorithm combined with the MDL by far.} 
        	
	        % 	\item It is proposed to combine the predictions of all NN models into an angle spectrum, whereupon a peak detection algorithm is devised and applied to find the AOA estimates. 
        \end{enumerate}
        
        The following notations apply throughout the entire paper. The transpose operator is denoted by $( \boldsymbol{\cdot})^T$, $( \boldsymbol{\cdot})^H$ stands for complex conjugate transpose and $E[\boldsymbol{\cdot}]$ is the expectation operator. Scalars are denoted as $a$ or $A$ (lightface), whereas $\mathbf{a}$ (boldface lowercase) denotes a column vector and $\mathbf{A}$ (boldface uppercase) is a matrix. $\Re(\boldsymbol\cdot)$ and $\Im(\boldsymbol\cdot)$ represent the real and imaginary part of a complex variable or function, respectively. The $n\times n$ identity matrix is denoted as $\mathbf{I}_n$ and $\mathrm{diag}(\mathbf{a})$ is a diagonal matrix with the elements of $\mathbf{a}$ on the diagonal. 
        
    The remainder of this paper is structured as follows. The data model and the problem statement are discussed in Section~\ref{sec:data_model}. The proposed AOA estimator is presented in Section~\ref{sec:MLF}, whereupon performance metrics  are described in Section~\ref{sec:metrics}. The conducted simulations and their results are presented and analyzed in Section~\ref{sec:sim_results} followed by conclusions in Section~\ref{sec:conclusions}.
        
\section{Data Model and Problem Statement}
\label{sec:data_model}

    Let’s consider $Q$ narrowband sources (i.e., incident plane waves) in the far-field of a uniform linear array (ULA) composed of $N$ sensors with inter-element spacing $d$. It is assumed that the sources and the sensors are all in the same plane, such that the direction-of-arrival (DOA) of each incident plane wave can be described by a single parameter, i.e., an angle-of-arrival (AOA). Hence, a one-dimensional (1D) AOA estimation problem is considered. The AOA of the $q$\textsuperscript{th} wave equals $\theta_q$, with $q=1,\dots,Q$, and is defined with respect to the ULA's broadside. The problem addressed in this paper is the joint estimation of the source number $Q$ and the AOAs $\theta_1,\dots,\theta_Q$ given $T$ snapshots of the sensor array output.
    % as shown on Fig.~\ref{fig:data_model}. 
    
    % \begin{figure}[tb]
    % 	\centering
    % 	\def\svgwidth{0.8\columnwidth}
    % 	\input{Images/data_model_v2.pdf_tex}
    % 	\caption{ULA configuration for 1D AOA estimation.}
    % 	\label{fig:data_model}
    % \end{figure}
    
    The sensor array output $\mathbf{y}(t)\in\mathbb{C}^{N \times 1}$, sampled at time instance $t$, is represented by the signal model  
    \begin{equation}
    	\label{eq:data_model}
    	\mathbf{y}(t) = \mathbf{A}(\theta_1,\dots,\theta_Q)\mathbf{s}(t) + \mathbf{n}(t),
    \end{equation}
    where $\mathbf{s}(t)\in\mathbb{C}^{Q \times 1}$ and $\mathbf{n}(t)\in\mathbb{C}^{N \times 1}$ represent the signal waveforms and the element noise, respectively, and $\mathbf{A}(\theta_1,\dots,\theta_Q)  \in\mathbb{C}^{N \times Q}$ is the array manifold consisting of $Q$ steering vectors, i.e.
    \begin{equation}
        \mathbf{A}(\theta_1,\dots,\theta_Q)=[\mathbf{a}_1(\theta_1),\dots,\mathbf{a}_Q(\theta_Q)].
    \end{equation}
    The $q$\textsuperscript{th} steering vector $\mathbf{a}_q(\theta_q) \in\mathbb{C}^{N \times 1}$ describes the array response to the $q$\textsuperscript{th} wave and is defined as 
    \begin{equation}
    	\mathbf{a}_q(\theta_q) = \big[1,e^{j\frac{2\pi}{\lambda}d\sin \theta_q},\dots, e^{j\frac{2\pi}{\lambda} (N-1)d\sin \theta_q}\big]^T ,
    \end{equation}
    where $\lambda$ is the wavelength of the transmitted signal.
    
    In this paper, $\mathbf{s}(t)$ and $\mathbf{n}(t)$ are both assumed to be i.i.d. zero-mean complex Gaussian random variables. Hence, the signal covariance matrix is given by 
    \begin{equation}
    \label{eq:sig_cov_mat}
        \mathbf{P}=E[\mathbf{s}(t)\mathbf{s}^H(t)] =\mathrm{diag}([\sigma_1^2,\dots,\sigma_Q^2]^T),
   \end{equation}
    where $\sigma_q^2$ denotes the variance of the $q$\textsuperscript{th} signal. It is assumed that the noise power is equal over all sensors, such that the noise covariance matrix is defined as 
    \begin{equation}
        \mathbf{Q}=E[\mathbf{n}(t)\mathbf{n}^H(t)]=\nu^2\mathbf{I}_N,
   \end{equation}
    where $\nu^2$ is the noise variance. Hence, the covariance matrix equals 
    \begin{equation}
        \mathbf{R}=E[\mathbf{y}(t)\mathbf{y}^H(t)]=\mathbf{APA}^H+\mathbf{Q}.
   \end{equation}
    In practice, $\mathbf{R}$ has to be estimated from noisy array measurements. For an array measurement consisting of $T$ snapshots, the maximum likelihood estimate, $\mathbf{\hat{R}}$, is computed as
    \begin{equation}
    	\label{eq:Rhat}
    	\hat{\mathbf{R}} = \frac{1}{T} \sum_{t=1}^T \mathbf{y}(t) \mathbf{y}^H(t),
    \end{equation}
    where it is assumed that the AOAs $\theta_1,\dots,\theta_Q$ (and therefore $Q$ as well) are identical for all $T$ snapshots $\{\mathbf{y}(1),\dots,\mathbf{y}(T)\}$. 
    
    The machine learning framework developed for the joint AOA estimation problem is presented in the next section.

\section{Supervised-Learning-Based Joint AOA Estimation Framework}
\label{sec:MLF}

    The proposed learning-based estimator comprises two main components: (I) an ensemble of learning-based classifiers, organized along a framework, and (II) a procedure to convert the predictions of these classifiers to angle-of-arrival (AOA) estimates. We proceed by first presenting each component and the related aspects, followed by a description of the deployment procedure of the estimator as a whole. 

    \subsection{AOA Estimation Framework}
    \label{subsec:framework}
        AOA estimation in scenarios with a variable number of sources implies that the number of parameters to be estimated is variable too. As this is not straightforward to implement using existing supervised learning algorithms, a framework is devised to recast the problem. This framework is the core of the estimator as it defines the number of classifiers in the ensemble, what their target outputs should be during training, and how their predictions should be interpreted and converted into AOA estimates during deployment. 
        
        \subsubsection{Multi-Source AOA Estimation Through Classification}
        \label{subsubsec:AOA_classification}
            Consider the array's field of view (FOV) defined by the interval $[\theta_\mathrm{min}, \theta_\mathrm{max})$. This interval is discretised into $M$ non-overlapping segments. Although not necessary, the presented method is specialized to a regular discretisation. Therefore, each segment spans $\Delta\theta$ degrees, where
        	\begin{equation}
        	\label{eq:del_theta}
        		\Delta\theta=\frac{\theta_\mathrm{max} - \theta_\mathrm{min}}{M}.
        	\end{equation}
        	Hence, $\Delta\theta$ denotes the angle resolution of the FOV discretisation, henceforth abbreviated as the FOV resolution. \begin{subequations}\label{eq:segment_bounds} The $i$\textsuperscript{th} FOV segment is defined by the interval $[\theta_{i,\mathrm{min}}, \theta_{i,\mathrm{max}})$, where 
            \begin{align}
                \theta_{i,\mathrm{min}} &= \theta_\mathrm{min}+(i-1)\Delta\theta \\
                \theta_{i,\mathrm{max}} &= \theta_\mathrm{min}+i\Delta\theta 
            \end{align}
            \end{subequations}
        	and $i=1,\dots,M$. Using the discretised FOV, we recast the AOA estimation problem as a classification problem: the proposed estimator aims to find those, and only those, FOV segments which include at least one of the AOAs $\theta_1,\dots,\theta_Q$. This is a so-called multi-label multi-class (or simply multi-label) classification problem \cite{zhang2013review}: $M$ distinct labels (here, non-overlapping FOV segments) exist, of which at most\footnote{The number of labels to be assigned is smaller than $Q$ if multiple AOAs belong to the same FOV segment.} $Q$ should be assigned to a single instance (here, a collection of $T$ snapshots of the array output). 
        	
        	Multi-label classification problems have been addressed successfully by transforming them into multiple single-label classification problems through the random $k$-labelsets (RA$k$EL) method \cite{tsoumakas2010random}. This method is the basis for the AOA estimation framework, hence we present its main principles below. 

        % In multi-label classification problems a single instance can be assigned a set of labels instead of just a single one .
    % 	Multi-label classification problems can be approached by either applying a multi-label learning algorithm directly, or by transforming it using a (problem) transformation algorithm \cite{tsoumakas2010random}. The AOA estimator proposed in this work is based on the  
    	
    	\subsubsection{RAkEL for Multi-Label Classification \cite{tsoumakas2010random}}
    	\label{subsubsec:RAkEL_default}
            RA$k$EL transforms a multi-label problem of $M$ labels, $\{\lambda_1,\dots,\lambda_M\}$, into $m$ single-label problems of $2^k$ labels (where $k<M$) such that it can be solved by $m$ single-label classifiers $h_1,\dots,h_m$. This is achieved in two steps. First, the multi-label problem is divided in $m$ smaller (but still multi-label) problems by generating $m$ subsets of $k$ labels (called $k$-labelsets). The second step is the transformation of the smaller multi-label problems into single-label problems via a method called label powerset (LP). The LP of $k$-labelset $R_j$ ($j=1,\dots,M$), denoted as $\mathcal{P}(R_j)$, is the set containing all $2^k$ possible subsets of $R_j$ as its elements. For example, if $k=2$ and $R_j=\{\lambda_a,\lambda_b\}$, then $\mathcal{P}(R_j)=\{\{\}, \{\lambda_a\}, \{\lambda_b\}, \{\lambda_a,\lambda_b\}\}$. Hence, by defining $2^k$ new labels, each of them representing a different element of $\mathcal{P}(R_j)$, the $j$\textsuperscript{th} multi-label problem can be solved indirectly by single-label classifier $h_j$ by selecting 1 out of these $2^k$ labels.
            % A distinction is made between RA$k$EL\textsubscript{d} and RA$k$EL\textsubscript{o}, where the subscripts d and o stand for the disjoint and overlapping versions, respectively. With the disjoint (d) version of RA$k$EL, the labelsets are generated by random sampling without replacement, such that each of the $M$ labels is included in exactly one of the $k$-labelsets. 
                        
            The $k$-labelsets can be generated either via random sampling with or without replacement, referred to as RA$k$EL\textsubscript{o} and RA$k$EL\textsubscript{d}, respectively. Here, the subscript `o' stands for overlapping and the `d' for disjoint. With RA$k$EL\textsubscript{o}, a label could be included in multiple $k$-labelsets, in which case the final prediction on whether to assign this label is obtained by a majority voting procedure. When applying RA$k$EL\textsubscript{o}, it is recommended \cite{tsoumakas2010random} to use a small $k$ ($k=3$ is given as an example) and $M < m < 2M$, as it is more efficient to use a large $m$ than a large $k$ in terms of computational burden. It is shown in \cite{tsoumakas2010random} that, averaged over 8 datasets from different fields, RA$k$EL\textsubscript{o} outperforms RA$k$EL\textsubscript{d} in terms of the $F_1$-score, a measure for predictive performance. 
    	   
    	\subsubsection{Combining RAkEL\textsubscript{d} and RAkEL\textsubscript{o} for AOA Estimation}
    	\label{subsubsec:RAkEL_comb}
    	    In this paper, RA$k$EL is applied for the sake of joint source number and AOA estimation. Hence, the labels $\lambda_1,\dots,\lambda_M$ represent the FOV segments, where the $i$\textsuperscript{th} segment is defined by the interval $[\theta_{i,\mathrm{min}}, \theta_{i,\mathrm{max}})$ \eqref{eq:segment_bounds}. However, rather than using either RA$k$EL\textsubscript{o} or RA$k$EL\textsubscript{d}, we propose to combine both variants, because of the following. When generating the $k$-labelsets via random sampling with replacement in RA$k$EL\textsubscript{o}, one cannot control the number of $k$-labelsets in which a particular label is included. More specifically, as each label is selected with equal probability, there is a probability of $((M-k)/M)^m$ for a label not to be included in any $k$-labelset. For the application addressed in this work, this implies that certain segments of the FOV might not be considered by the AOA estimator. This is clearly problematic as the estimator would not be able to `see' waves with AOAs within those segments. 
    	   % This probability can be decreased by increasing $m$ or $k$, but this also further increases the computational complexity.
    	    To circumvent this problem without having to increase $m$ and/or $k$ (which increases the computational burden), it is proposed to approximate RA$k$EL\textsubscript{o} by using $L$ independent `layers' of RA$k$EL\textsubscript{d}. Consequently, each label is included in exactly $L$ $k$-labelsets and the majority voting procedure of RA$k$EL\textsubscript{o} can be applied for all labels $\lambda_1$,\dots,$\lambda_M$. The total number of classifiers in this layered framework equals 
        	\begin{equation}
        		\label{eq:num_classifiers}
        		m = L\lceil M/k\rceil,
        	\end{equation}
        	where $\lceil \boldsymbol \cdot \rceil$ rounds up the argument to the nearest integer\footnote{By proper choice of $M$, the existence of a labelset consisting of less than $k$ labels can be prevented and rounding can be discarded.}. It is worthwhile to note that both increasing the FOV resolution (i.e., decreasing $\Delta\theta$) and increasing the number of layers $L$ results in a larger number of classifiers in the framework. An example of the proposed layered framework is presented in the first 2 columns of Table~\ref{tab:example}.

        	\setlength{\tabcolsep}{3pt}
        	\setlength\aboverulesep{2pt}
            \setlength\belowrulesep{2pt}
        	\begin{table}[tbp]
        		\centering	
        		\caption{Example RA$k$EL-based \cite{tsoumakas2010random} AOA estimation framework with $k=2$, $M=4$, $L=2$.}
        		\label{tab:example}
        		\begin{tabularx}{\columnwidth}{@{}c cXccccc }
        			\toprule
        			
        			classif. & $k$-labelset & \multicolumn{1}{c}{label-subset} & \multicolumn{1}{c}{predict.} &\multicolumn{4}{c}{$\bar{P}_{i,j, \tilde{k}}$ (\ref{eq:prob_pred})} \\ \cmidrule{5-8}
        			
        			$h_j$ & $R_j$ & \multicolumn{1}{c}{$\tilde{R}_{j,\tilde{k}}$} & $\tilde{P}_{j,\tilde{k}}$ & $\lambda_1$ & $\lambda_2$ & $\lambda_3$ & $\lambda_4$ \\ \midrule

        			\multicolumn{8}{@{}l}{\textbf{Layer 1}} \\ \cmidrule{3-8}
        			
        			\multirow{4}{*}{$h_1$} & \multirow{4}{*}{$\{\lambda_1,\lambda_3\}$}& $\tilde{R}_{1,1}=\{\}$ & $\tilde{P}_{1,1}$ & \cellcolor{gray!25}0 & 0 & \cellcolor{gray!25}0 & 0 \\
        			
        			& & $\tilde{R}_{1,2}=\{\lambda_1\}$ & $\tilde{P}_{1,2}$& \cellcolor{gray!25}$\tilde{P}_{1,2}$ & \multirow{2}{*}{$\rvdots$} & \cellcolor{gray!25}0  & \multirow{2}{*}{$\rvdots$} \\
        			
        			& & $\tilde{R}_{1,3}=\{\lambda_3\}$ & $\tilde{P}_{1,3}$ & \cellcolor{gray!25}0 &  & \cellcolor{gray!25}$\tilde{P}_{1,3}$ &  \\
        			
        			& & $\tilde{R}_{1,4}=\{\lambda_1,\lambda_3\}$ & $\tilde{P}_{1,4}$ & \cellcolor{gray!25}$\tilde{P}_{1,4}$ & 0  & \cellcolor{gray!25}$\tilde{P}_{1,4}$  & 0 \\  \cmidrule{3-8}

        			\multirow{4}{*}{$h_2$} & \multirow{4}{*}{$\{\lambda_2,\lambda_4\}$} & $\tilde{R}_{2,1}=\{\}$ & $\tilde{P}_{2,1}$ & 0 & \cellcolor{gray!25}0 & 0 & \cellcolor{gray!25}0 \\ 
        			
    	    		& & $\tilde{R}_{2,2}=\{\lambda_2\}$ & $\tilde{P}_{2,2}$ & \multirow{2}{*}{$\rvdots$} & \cellcolor{gray!25} $\tilde{P}_{2,2}$  & \multirow{2}{*}{$\rvdots$}  & \cellcolor{gray!25}0 \\
    	    		
        			& & $\tilde{R}_{2,3}=\{\lambda_4\}$  & $\tilde{P}_{2,3}$ &  &\cellcolor{gray!25} 0  &   & \cellcolor{gray!25}$\tilde{P}_{2,3}$ \\
        			
        			& & $\tilde{R}_{2,4}=\{\lambda_2,\lambda_4\}$ & $\tilde{P}_{2,4}$ & 0 & \cellcolor{gray!25}$\tilde{P}_{2,4}$  & 0 & \cellcolor{gray!25}$\tilde{P}_{2,4}$ \\ \midrule
        			
        			\multicolumn{8}{@{}l}{\textbf{Layer 2}} \\ \cmidrule{3-8}

        			$h_3$ &$\{\lambda_1,\lambda_4\}$ & \multicolumn{1}{c}{$\rvdots$}  & $\rvdots$ & \cellcolor{gray!25}$\rvdots$ & $\mathbf{0}$ & $\mathbf{0}$ & \cellcolor{gray!25} $\rvdots$ \\ 
        			
        			$h_4$ & $\{\lambda_2,\lambda_3\}$ & \multicolumn{1}{c}{$\rvdots$}  & $\rvdots$ & $\mathbf{0}$ & \cellcolor{gray!25}$\rvdots$ & \cellcolor{gray!25}$\rvdots$ & $\mathbf{0}$  \\
        			\bottomrule
        		\end{tabularx}{}
        	\end{table} 
        	
    \subsection{Converting Classifier Predictions to joint AOA Estimates}
    \label{subsec:estimates}
    
        Section~\ref{subsec:framework} described how the AOA estimation problem is decomposed into multiple single-label classification problems. Here, we present how the classifiers' predictions are converted to AOA and source number estimates when the estimator is deployed. To be as generic as possible regarding the learning algorithm, it is assumed that the single-label learning algorithms' prediction comprises a set of probabilities, rather than a single index.
        
    	\subsubsection{Classifier Predictions}
        \label{subsubsec:classif_pred}	
        
            Let's denote the elements of $\mathcal{P}(R_j)$ (i.e., the label subsets of $R_j$) as $\tilde{R}_{j,1},\dots,\tilde{R}_{j,2^k}$. Hence, by definition it holds that $\tilde{R}_{j,\tilde{k}} \subseteq R_j \subset \{\lambda_1,\dots,\lambda_M\}$, where $\tilde{k}=1,\dots,2^k$. Furthermore, we denote any prediction of classifier $h_j$ as the set  $\{\tilde{P}_{j,1},\dots,\tilde{P}_{j,2^k}\}$, for which it holds that $0\leq \tilde{P}_{j,\tilde{k}} \leq{1}$ ($\tilde{k}=1,\dots,2^k$) and that  $\tilde{P}_{j,1}+\tilde{P}_{j,2}+\dots+\tilde{P}_{j,2^k}=1$. Thus, $\tilde{P}_{j,\tilde{k}}$ is directly related to the label subset $\tilde{R}_{j,\tilde{k}}$, as visualised by columns 3 and 4 of Table~\ref{tab:example}. It represents the probability that there is at least one AOA within every FOV segment represented by the labels in $\tilde{R}_{j,\tilde{k}}$, according to classifier $h_j$.
        	
            Rather than converting the probabilistic predictions $\tilde{P}_{j,1},\dots,\tilde{P}_{j,2^k}$ to Boolean variables (i.e., 1 for the highest probability and 0 for all the others) and subsequently applying the majority voting procedure of RA$k$EL\textsubscript{o}, we adopt another approach to estimate the AOAs. This approach prevents the loss of information in this stage of the estimation process and it prevents a single wave with an AOA close to the border between two neighbouring FOV segments to result in a double AOA estimate\footnote{Also, the proposed method accommodates the use of different FOV discretisations (e.g., random non-uniform discretisations) for the different framework layers. A first step in this direction is presented in \cite{kanters2020direction}.}. 
        
        \subsubsection{Computing Source Number and AOA Estimates}
        \label{subsubsec:estimation}
            
\begin{figure}[tb]
        		\centering
        		\def\svgwidth{\columnwidth}
        		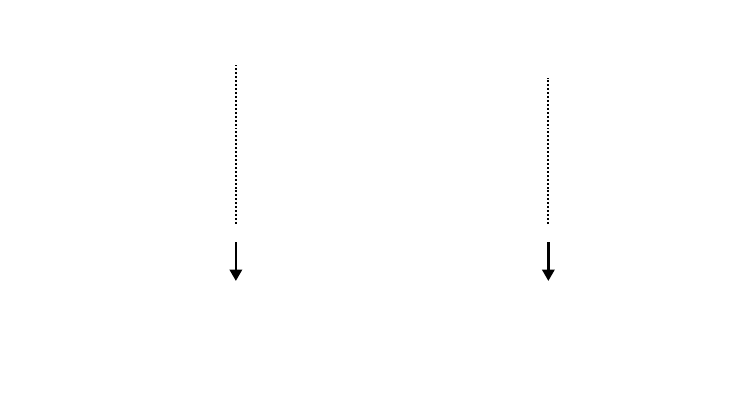
        		\vspace*{-1.3cm}
        		\caption{Example spectrum and resulting AOA estimates.}
        		\label{fig:spectrum}
        	\end{figure}

            First, all probabilistic predictions $\tilde{P}_{j,\tilde{k}}$ (with $j=1,\dots,m$ and $\tilde{k}=1,\dots,2^k$) are converted to per-label-predictions $\bar{P}_{i,j,\tilde{k}}$ ($i=1,\dots,M$) according to 
        	\begin{equation}
        	    \label{eq:prob_pred}
        	    \bar{P}_{i,j, \tilde{k}} = 
        	    \begin{cases}
                    \tilde{P}_{j,\tilde{k}}, & \text{ $\lambda_i \in \tilde{R}_{j,\tilde{k}}$},\\
                    0, & \text{otherwise}.
                \end{cases} 
        	\end{equation}
            An example is presented in the 4 rightmost columns of Table~\ref{tab:example}. Then, the per-label-predictions $\bar{P}_{i,j,\tilde{k}}$ are combined into segment probabilities $P_1,\dots,P_M$ as
    	    \begin{equation}
        	    \label{eq:grid_prob}
        	    P_i = \frac{1}{L} \sum_{j=1}^{m} \sum_{\tilde{k}=1}^{2^k} \bar{P}_{i,j, \tilde{k}}.
        	\end{equation}
            The division by $L$ in (\ref{eq:grid_prob}) guarantees that $0 \leq P_i \leq 1$, as each label is included in exactly $L$ $k$-labelsets. Hence, $P_i$ represents the probability that there is at least one AOA within the $i$\textsuperscript{th} FOV segment, according to the $L$ classifiers evaluating it. Finally, we interpret the sequence of probabilities $P_1,\dots,P_M$ as an angle spectrum, similar to the work presented in \cite{liu2018direction}. In order to jointly extract the source number and the AOAs from this spectrum, we propose to use a straightforward peak detection algorithm. This algorithm locates all spectrum peaks above a threshold and returns the number of peaks as the source number estimate, $\hat{Q}$, and their arguments as the AOA estimates $\hat{\theta},\dots,\hat{\theta}_{\hat{Q}}$. Since each peak has a plateau width of $\Delta\theta$ \eqref{eq:del_theta} degrees (the resolution of the discretised FOV), the centre of the plateau is taken as the estimate. The possible AOA estimates are therefore defined by the centres of the FOV segments $c_1,\dots,c_M$, where
    	    \begin{equation}
       	        c_i=\frac{1}{2}(\theta_{i,\mathrm{min}}+\theta_{i,\mathrm{max}})=\theta_\mathrm{min}+(i-1/2)\Delta\theta.  
        	    \end{equation}
            An example spectrum and its corresponding AOA estimates are presented in Fig.~\ref{fig:spectrum}.

        \begin{figure*}[t!]
    		\centering
    		\includegraphics[width=\textwidth]{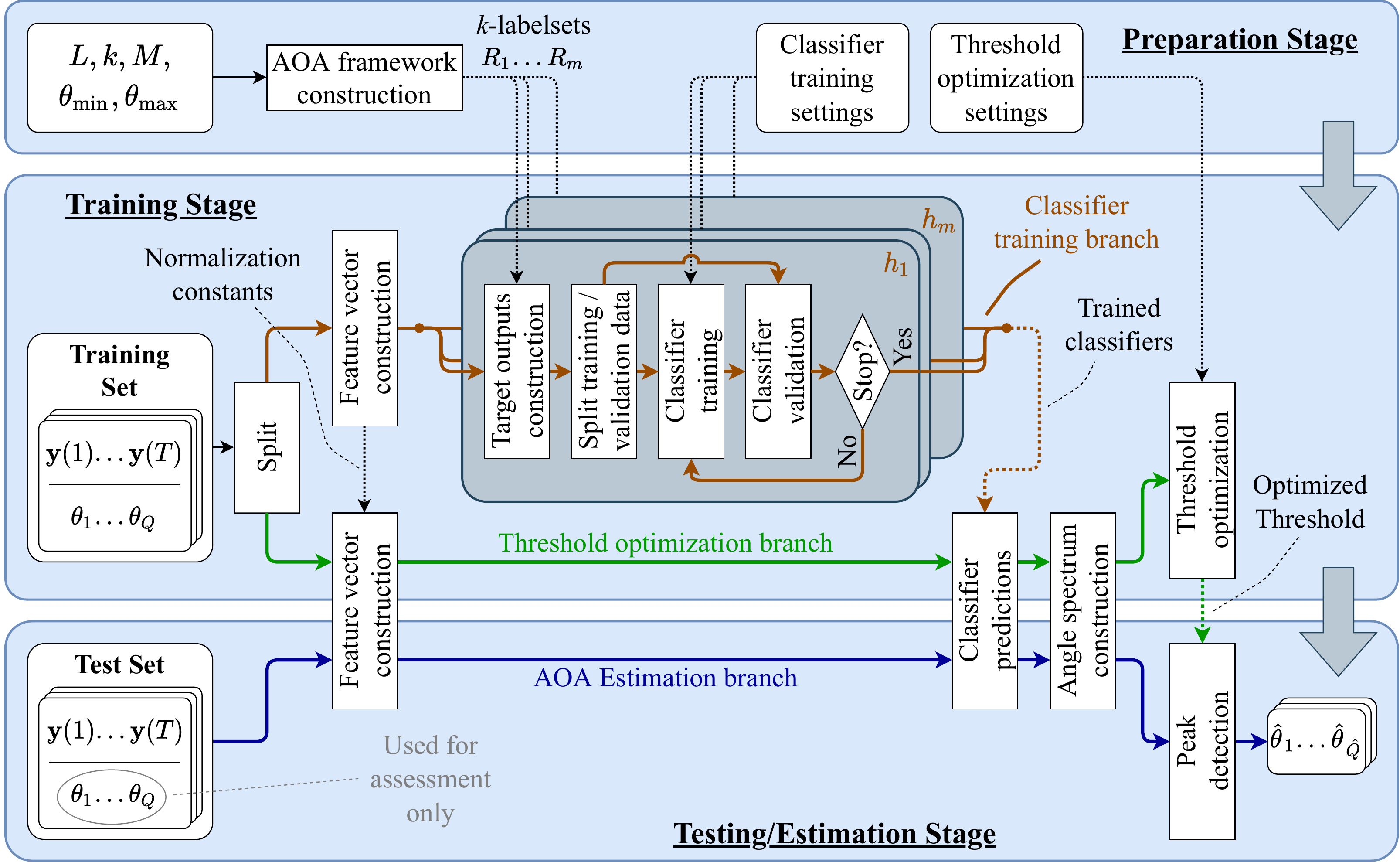}
    		\caption{Flowchart of AOA estimator deployment procedure}
    		\label{fig:flowchart}
    	\end{figure*}
    	
    \subsection{Deployment Process Flow}
    \label{subsec:train_test}

        The complete AOA estimation procedure is visualised in Fig.~\ref{fig:flowchart}. Three stages can be identified: (I) the preparation stage, (II) the training stage and (III) testing/estimation stage. Details regarding each of these stages are explained next. 

    	\subsubsection{Preparation Stage}
    	\label{subsubsec:prep_stage}
    	    The core of the preparation stage is the construction of the RA$k$EL-based framework, i.e., the generation of the $k$-labelsets $R_1,\dots R_m$, as described in Section~\ref{subsec:framework}. For this, the array's FOV and the framework's topology need to be defined through the parameters $\theta_{\mathrm{min}}$, $\theta_{\mathrm{min}}$, and $L$, $k$, $M$, respectively, whereupon the FOV resolution $\Delta\theta$ (\ref{eq:del_theta}) and the number of classifiers $m$ (\ref{eq:num_classifiers}) follow automatically. 
    	    
    	    Besides the framework construction, a number of settings regarding the classifier training (e.g., the learning algorithm and its corresponding design parameters) and the threshold optimization need to be defined during the preparation stage as well. Details are clarified below.   
    	   % Since the former depends on the chosen learning algorithm, no further discussion is presented here. The required threshold optimization settings are clarified below. 

    	\subsubsection{Training Stage}
    	\label{subsubsec:training_stage}
    	    
    	    In the training stage, the AOA estimator is optimized based on training data. We assume a training set of $D_{\mathrm{trn}}$ instances is available, where an instance contains $T$ snapshots of the array output (\ref{eq:data_model}) paired with the corresponding AOAs (i.e., the AOAs for which these array outputs were computed). However, as the training stage is composed of two branches, (I) the classifier training branch and (II) the threshold optimization branch, the training set must be split in two (not necessarily equally large) parts.

    	    The details of the classifier training branch depend on the employed learning algorithm. However, in general, the procedure contains the following steps. First, the training data need to be prepared such that they can be used for supervised learning, meaning input-output pairs need to be composed. The input component of an input-output pair, the so-called feature vector, contains the available information based on which the learning algorithm computes its prediction. Hence, in the present work, the feature vector is derived from the array data. It is worthwhile to note that every instance from the training set (the part used for classifier training) yields $m$ input-output pairs, i.e., one for each classifier, all sharing the same feature vector. After computing the feature vectors for all instances, element-wise feature normalization is applied, since some learning algorithms are sensitive to scale \cite{zheng2018feature}.
    	   % This form of normalization, often called standardization, simplifies the learning process as the NNs do not have to adapt to features (numerical values) of different ranges \cite{francois2017deep}. 
    	    The output components of input-output pairs represent the prediction targets. Contrary to the inputs, they need to be computed for classifier (and each instance, clearly) individually, as each classifier is associated with its own $k$-labelset. Since it is assumed that each prediction of a single-label classifier comprises $2^k$  probabilities (Section~\ref{subsubsec:classif_pred}), this must also apply to the prediction targets. Hence, for one particular training instance, the targets for classifier $h_j$ ($j=1,\dots,m$), denoted as $\{{\tilde{P}^{(t)}_{j,1}},\dots,{\tilde{P}^{(t)}_{j,2^k}} \}$, are computed as
        	\begin{equation}
        		\label{eq:des_class_pred}
        		\tilde{P}^{(t)}_{j,\tilde{k}} = 
        		\begin{cases}
        			1, & \text{if } \tilde{R}_{j,\tilde{k}}=(R_j \cap \bar{\Lambda}) \\
        			0, & \text{otherwise},
        		\end{cases}
        	\end{equation}
        	where $\tilde{k}=1,\dots,2^k$ and $R_j$ is the $k$-labelset associated with $h_j$, with label-subsets $\tilde{R}_{j,1},\dots,\tilde{R}_{j,2^k}$. In \eqref{eq:des_class_pred}, $\bar{\Lambda}$ is the set containing exactly those labels representing FOV segments which include at least one of the instance's AOAs. Hence, it is defined as 
        	\begin{equation}
        	\label{eq:lambda_bar}
        	\begin{split}
	        	    \bar{\Lambda} = \{\lambda_i \, | \, & i\in\{1,\dots,M\} \, \wedge \\
	        	    & (\exists \theta_q)[\theta_q\in \Theta \wedge \theta_{i,\mathrm{min}} \leq \theta_q < \theta_{i,\mathrm{max}}]  \},
        	\end{split}
        	\end{equation}
            where $\Theta=\{\theta_{1},\dots,\theta_{Q}\}$ with $\theta_{1},\dots,\theta_{Q}$ and $Q$ being the true AOAs and the true source number of the instance under consideration, respectively. After composing the input-output pairs for all instances and all classifiers, the actual training is carried out. As each classifier learns its own mapping, it is proposed to track the learning progress of each classifier individually by means of a validation set in order to determine when to stop training. 
            
            Once the training has been terminated for all classifiers, the threshold optimization branch is initiated. This branch aims to optimize the threshold level (i.e., probability level) employed in the peak detection algorithm (Fig.~\ref{fig:spectrum}). The process is as follows. First, feature vectors are computed for all threshold optimization training instances.
            % \footnote{These instances are not used for classifier training/validation to prevent the threshold from being optimized on unrealistically accurate classifier predictions.}.
            This is done in the same way as in the classifier training branch, except that the feature-wise normalization is done using the normalization constants (feature-wise means and variances) derived from the classifier training data. In this way, we emulate the estimation stage, in which one can only normalize based on training data as well. The feature vectors are fed through the ensemble of trained classifiers, whereupon the resulting predictions $\tilde{P}_{j,\tilde{k}}$ are converted to angle spectra according to the procedure described in Section~\ref{subsec:estimates}. As these spectra (of which there are as many as there are threshold optimization training instances) only contain values between 0 and 1 by definition, the optimal threshold must be between these values as well. The actual threshold optimization is a matter of computing the AOA and source number estimates for all spectra for a set of threshold values (to be defined in the preparation stage). The threshold level that maximizes the number of spectra for which the estimated source number $\hat{Q}$ equals the true source number $Q$ is considered optimal and is used within the estimation stage. 
    	
    	\subsubsection{Testing/Estimation Stage}
    	\label{subsubsec:testing_stage}
            After finishing the training stage, the estimator can be applied for AOA estimation. For each instance, the estimation procedure is similar to the one described by the threshold optimization branch, with the only difference being that the optimal threshold level is now known and can be applied directly. To assess the performance of the estimator, a test set of $D_{\mathrm{tst}}$ instances is used.
            
\section{Performance Metrics}
\label{sec:metrics}	
    	
	In the present work, the accuracy of the estimates obtained from the proposed angle-of-arrival (AOA) estimator depends on (I) the framework topology (defined by framework parameters $\theta_\mathrm{min}$, $\theta_\mathrm{max}$ and $L$, $k$, $M$), and (II) the predictive performance of the single-label classifiers used within the framework. 
	The metrics employed to study the impact of the above on the AOA estimates are defined below. 
	
    \subsection{RMSE and $P(\hat{Q}=Q)$}
    \label{subsubsec:RMSE}	
        
       The accuracy of the AOA estimates is evaluated by means of the root-mean-square error (RMSE), which is computed as
        \begin{equation}
            \label{eq:RMSE}
            \mathrm{RMSE} = \sqrt{\frac{1}{P'}\sum_{p'=1}^{P'} \bigg[\frac{1}{Q_{p'}} \sum_{q=1}^{Q_{p'}} (\theta_{{p'},q}-\hat{\theta}_{{p'},q})^2 \bigg]},
        \end{equation}
        where $\theta_{p',q}$ and $\hat{\theta}_{p',q}$ are the $q$\textsuperscript{th} true AOA and the $q$\textsuperscript{th} AOA estimate in (test) instance $p'$, respectively, $Q_{p'}$ is the number of true AOAs in instance $p'$ and $P'$ is the number of evaluated instances. For each instance, the AOAs and AOA estimates are sorted in the same order before computing the RMSE. 
        
        It can be seen that only instances for which the source number estimate $\hat{Q}_{p'}$ equals the true source number $Q_{p'}$ can be included in the RMSE computation. As there might be instances for which this does not apply, an additional metric $P(\hat{Q}=Q)$, representing the probability that the source number estimate is correct, is defined as
        \begin{equation}
            \label{eq:PQQhat}
            P(\hat{Q}=Q) = \frac{P'}{P} \times 100\%,
        \end{equation}
        where
        \begin{equation}
            P' = \mathrm{num}(\hat{Q}_{p} = Q_{p}).
        \end{equation}
        Here, $Q_{p}$ and $\hat{Q}_{p}$ are the true and estimated source number for instance $p$, respectively, $p=1,\dots,P$ where $P$ is the total number of evaluated instances (hence, $P=D_{\mathrm{tst}}$) and $\mathrm{num}(x)$ denotes the operation of counting the number of statements for which $x$ is true. 

        % \begin{equation}
        %     P=num(\hat{Q}_{p'} = Q_{p'}). 
        % \end{equation} 
        
        % represents the total number of instances for which the source number estimate equals the true source number and $q=1,\dots,Q_{p'}$ with $Q_{p'}$ being the true source number in instance $p'$. 
        
        % Therefore, an additional metric $P(\hat{Q}=Q)$, representing the probability that the source number estimate is correct, is defined as
        % \begin{equation}
        %     \label{eq:PQQhat}
        %     P(\hat{Q}=Q) = \frac{num(\hat{Q}_p = Q_p)}{P} \times 100\%,
        % \end{equation}
        % where $Q_p$ and $\hat{Q}_p$ are the true and the estimated source number for the $p$\textsuperscript{th} instance, respectively, $p=1,\dots,P$ and $P$ is the total number of evaluated instances to which the estimator is applied (i.e., $P=D_{\mathrm{tst}}$).
        
    \subsection{Joint AOA Estimation Success Rate}
    \label{subsubsec:combined2}
        Given that the proposed AOA estimator performs jointly the source number detection and the AOA estimation, a metric is devised which takes into account both these aspects. It is based on the $success~rate$ proposed in \cite{zhu2020two} and expressed as		
        \begin{equation}
            \label{eq:fsr}
            f_\mathrm{sr}(\tilde{\theta}) = \frac{\mathrm{num}\Big(\big[\hat{Q}_p=Q_p\big] \cap \big[|\hat{\theta}_{p,q} - \theta_{p,q}|\leq \tilde{\theta} {\,} \big]\Big)}{P} \times 100\%,
        \end{equation}
        where $q=1,\dots,Q_p$ and $p=1,\dots,P$, with $Q_p$ and $P$ as defined above. Hence, \eqref{eq:fsr} implies that the joint AOA estimate for the $p$\textsuperscript{th} instance is successful only if the source number is estimated correctly, i.e., $\hat{Q}_p=Q_p$, and all AOA estimation errors $|\hat{\theta}_{p,q} - \theta_{p,q}|$ (computed after sorting) are smaller than or equal to the maximum allowed AOA estimation error $\tilde{\theta}$. It is worthwhile to note that estimation errors up to $\Delta\theta/2$ are expected due to the finite FOV resolution.
        
        As a reference for the success rate $f_\mathrm{sr}(\tilde{\theta})$, we introduce $f_{\mathrm{sr,exp}}(\tilde{\theta})$, which represents the success rate that would be expected if all classifiers in the framework were ideal, i.e., if their predictions $\tilde{P}_{j,\tilde{k}}$ equal the prediction targets $\tilde{P}^{(t)}_{j,\tilde{k}}$ \eqref{eq:des_class_pred} for all considered instances. Hence, $f_{\mathrm{sr,exp}}(\tilde{\theta})$ is a measure for success rate limitations imposed by the framework's topology. In the case of a regular FOV discretisation and uniformly distributed random AOAs sharing the interval $[\theta_\mathrm{min}, \theta_\mathrm{max})$, $f_{\mathrm{sr,exp}}(\tilde{\theta})$ is computed as
        \begin{equation}
        \label{eq:success_rate}
            f_{\mathrm{sr,exp}}(\tilde{\theta})=
            \begin{cases}
            \Big(\frac{\tilde{\theta}}{\Delta\theta/2}\Big)^Q f_{\mathrm{sr,exp},\mathrm{max}}, & \text{if } \tilde{\theta} < \Delta\theta/2, \\
             f_{\mathrm{sr,exp,max}}, & \text{otherwise}
            \end{cases},
        \end{equation}
        where
        \begin{equation}
        \label{eq:SR_pdf}
            f_{\mathrm{sr,exp,max}} = {M + 1 - Q \choose Q} \frac{Q!}{M^Q} \times 100\%.
        \end{equation}
        Derivations of \eqref{eq:success_rate} and \eqref{eq:SR_pdf} are presented in Appendix~\ref{sec:App_A} and \ref{sec:App_B}, respectively. It is worthwhile to note that (\ref{eq:success_rate}) assumes that the source number $Q$ is equal in all evaluated instances. If not, $f_{\mathrm{sr,exp}}(\tilde{\theta})$ is computed for all possible values of $Q$ individually and a (weighted) average is applied afterwards. 
        
    \subsection{$F_1$-score}	
    \label{subsubsec:base_perf}		
        Besides evaluating the source number and AOA estimates directly, the predictions of the single-label classifiers are evaluated as well. This is done by means of the $F_1$-score (see, e.g., \cite{tsoumakas2007random}). As the $F_1$-score is computed per label and per classifier, the notation $F_1(j,\tilde{k})$ is used from here, where the index $j=1,\dots,m$ refers to the classifier and the index $\tilde{k}=1,\dots,2^k$ to the label. The $F_1$-score is defined as the harmonic mean of two other metrics, precision and recall, with the subscript $1$ indicating that precision and recall both contribute with equal weights to the mean, i.e., 
        \begin{equation}\label{eq:F1}
            F_1(j, \tilde{k}) = 2\times \frac{\mathrm{precision}(j,\tilde{k}) \times \mathrm{recall}(j,\tilde{k}) }{\mathrm{precision}(j,\tilde{k})  + \mathrm{recall}(j,\tilde{k}) }. 
        \end{equation}
        Here, $\mathrm{precision}(j,\tilde{k})$ is defined as the ratio \begin{equation}
            \mathrm{precision}(j,\tilde{k}) = \frac{\mathrm{tp}_{j,\tilde{k}}}{\mathrm{tp}_{j,\tilde{k}}+\mathrm{fp}_{j,\tilde{k}}},
        \end{equation}
        where $\mathrm{tp}_{j,\tilde{k}}$ and $\mathrm{fp}_{j,\tilde{k}}$ denote the number of true and false positives (for label $\tilde{k}$ and classifier $h_j$), respectively\footnote{Since the classifiers' predictions are assumed to be probabilities rather than boolean variables, true/false positives/negatives are ill-defined. For the sake of $F_1$-score computation, we therefore assign boolean 1 to the label corresponding to the highest probability and boolean 0 to all the others.}. Hence, precision is a measure for a classifier's exactness. Furthermore, $\mathrm{recall}(j,\tilde{k})$ is defined as the ratio 
        \begin{equation}
            \mathrm{recall}(j,\tilde{k}) = \frac{\mathrm{tp}_{j,\tilde{k}}}{\mathrm{tp}_{j,\tilde{k}}+\mathrm{fn}_{j,\tilde{k}}},
        \end{equation}
        where $\mathrm{fn}_{j,\tilde{k}}$ denotes the number of false negatives (for label ${\tilde{k}}$ and classifier $h_j$). Hence, recall represents the fraction of all instances of label $\tilde{k}$ that are actually classified as such and is therefore a measure for a classifier's completeness. Consequently, it holds that $0\leq F_1(j,\tilde{k}) \leq 1$, with a higher value indicating a higher predictive performance.
        
        In this work, the assessment of all classifiers yields $m\times2^k$ $F_1$-scores. To assess these in a structured manner, we compute
        % This approach is tailored to scenarios with uniformly distributed random AOAs in combination with a regular discretisation of the array's FOV, and is defined as
        \begin{equation}
        \label{eq:F1_bar}
            \bar{F_1}(Q_h) = \frac{1}{m} \sum_{j=1}^m \frac{1}{|S_{j,Q_h}|} \sum_{\kappa\in S_{j,Q_h}} F_1(j,\kappa),
        \end{equation}
        where $F_1(j,\kappa)$ is computed according to \eqref{eq:F1} and where
        \begin{equation}
        S_{j,Q_h} = \{\tilde{k} \, | \, \tilde{k}\in \{1,\dots2^k\} \wedge |\tilde{R}_{j,\tilde{k}}| = Q_h\},     
        \end{equation}
        where $|S|$ denotes the cardinality of set $S$ and $Q_h=0,\dots,k$. In other words, $ S_{j,Q_h}$ is the set containing those indices $\tilde{k}$ that refer to the elements of $\mathcal{P}(R_j)$ (the label powerset of the $k$-labelset of classifier $h_j$) whose cardinality equals $Q_h$. For example, if $R_j=\{\lambda_a,\lambda_b\}$ ($k=2$) and we denote its subsets $\{\}, \{\lambda_a\}, \{\lambda_b\}, \{\lambda_a,\lambda_b\}$ as  $\tilde{R}_{j,1},\dots,\tilde{R}_{j,4}$, respectively, then $S_{j,0}=\{1\}$, $S_{j,1}=\{2,3\}$ and $S_{j,2}=\{4\}$. Hence, \eqref{eq:F1_bar} averages the $F_1$-scores of all labels representing label subsets with the same subset cardinality $Q_h$. It is important to note that if $\mathrm{tp}_{j,\tilde{k}} = 0$, then $F_1(j,\tilde{k})$ is not defined. In this case, this particular $F_1(j,\tilde{k})$ is excluded from the computation and the average is taken over all remaining valid $F_1$-scores. The latter can occur for various reasons, e.g., because $Q<k$, or simply because of the stochasticity of the AOAs and the $k$-labelsets.

\section{Simulations, Results and Analysis}
\label{sec:sim_results}

    In this section, we present the simulations that were conducted to assess the performance of the proposed angle-of-arrival (AOA) estimator and an analysis thereof. 
    
    \subsection{Simulation Set Up}
        
         A summary of the simulation parameters is presented in Table~\ref{tab:sim_params}. Details are given below.
        
    	\begin{table}[tbp]
    		\centering	
    		\caption{Simulation parameters}
    		\label{tab:sim_params}
    		\begin{tabularx}{\columnwidth}{@{} lXXX @{}}
    			\toprule[1pt]
    					& Parameter &\multicolumn{2}{c}{Value} \\ \cmidrule{3-4}
    					& & \multicolumn{1}{c}{Scenario I} & \multicolumn{1}{c}{Scenario II} \\ \midrule
    					
    			\multicolumn{4}{@{}l}{\textbf{Sources and Signals}}\\    
    			        & Source number &\multicolumn{1}{c}{\cellcolor{gray!25} $Q=2$} &\multicolumn{1}{c}{\cellcolor{gray!25}$Q\sim U(1,4)$} \\
        				& AOAs		&\multicolumn{2}{c}{$\theta_1,\dots,\theta_Q \sim U(-60^\circ, 60^\circ)$} \\
        				& SNR & \multicolumn{2}{c}{SNR $\in \{-10, 0, 10 \}$ dB} \\ 
        				 \midrule
       					
         		\multicolumn{4}{@{}l}{\textbf{AOA Estimation Framework}}\\
    					& FOV  & \multicolumn{2}{c}{$[\theta_\mathrm{min},\theta_\mathrm{max})=[-60^\circ, 60^\circ)$} \\
    					& \# framework layers & \multicolumn{2}{c}{$L\in\{1,3,5\}$} \\ 
    					& FOV resolution & \multicolumn{2}{c}{$\Delta\theta\in\{2^\circ,1^\circ\}$} \\
    					& Labelsets &\multicolumn{2}{c}{$k=3$} \\ \midrule
    					
    			\multicolumn{4}{@{}l}{\textbf{Sensor array}}\\
    					& Configuration      & \multicolumn{2}{c}{ULA} \\
    					& \# sensors & \multicolumn{2}{c}{$N=8$} \\ 
       					& Inter-element spacing & \multicolumn{2}{c}{$d=\lambda/2$} \\ \midrule

       			\multicolumn{4}{@{}l}{\textbf{Single-Label Classifiers}}\\ 
               			& Learning algorithm & \multicolumn{2}{c}{Feedforward neural networks}  \\
       					& Input layer, \# neurons & \multicolumn{2}{c}{$N^2=64$}  \\
       					& Hidden layers, \#   & \multicolumn{1}{c}{\cellcolor{gray!25}2} & \multicolumn{1}{c}{\cellcolor{gray!25}5}  \\
       					& Hidden layers, \# neurons & \multicolumn{1}{c}{\cellcolor{gray!25}64, 36} & \multicolumn{1}{c}{\cellcolor{gray!25}100, 100, 100, 100, 50} \\
       					& Hidden layers, activ. funct. & \multicolumn{2}{c}{ReLU}  \\
       					& Output layer, \# neurons & \multicolumn{2}{c}{$2^k=8$}  \\
      					& Output layer, activ. func.& \multicolumn{2}{c}{Softmax}  \\
    					& Optimizer & \multicolumn{2}{c}{Adam} \\
       					& \hspace{3pt} Learning rates  & \multicolumn{2}{c}{$\alpha=0.001$, $\beta_1=0.9$, $\beta_2=0.999$}  \\
       					& Loss function & \multicolumn{2}{c}{Categorical cross entropy} \\
   						& Mini-batch, \# instances & \multicolumn{2}{c}{$32$}  \\ \midrule 
   						
   	    			\multicolumn{4}{@{}l}{\textbf{Threshold Optimization}}\\
    					& Evaluated thresholds      & \multicolumn{2}{c}{$0.01,\,0.02,\,0.03, \dots,\, 1$}  \\ \midrule
   						
   	        		\multicolumn{4}{@{}l}{\textbf{Datasets}}\\ 					& \# snapshots per instance     &\multicolumn{2}{c}{$T=100$} \\			
    					& \# instances training set    & \multicolumn{1}{c}{\cellcolor{gray!25}$D_{\mathrm{trn}}=80\,000$} & \multicolumn{1}{c}{\cellcolor{gray!25}$D_{\mathrm{trn}}=320\,000$}  \\
        				& \hspace{3pt} Fraction classifier training &\multicolumn{2}{c}{$80\%$} \\
        				& \hspace{3pt} Fraction classifier validation &\multicolumn{2}{c}{$10\%$} \\
       					& \hspace{3pt} Fraction threshold optim. & \multicolumn{2}{c}{$10\%$}\\ 					
        				& \# instances test set & \multicolumn{2}{c}{$D_{\mathrm{tst}}=50\,000$}  \\  \midrule 
        				
   	        		\multicolumn{4}{@{}l}{\textbf{Benchmark Algorithms}}\\ 		
    					& AOA estimator    & \multicolumn{2}{c}{MUSIC} \\
        				& \hspace{3pt} Angle spectrum resolution &\multicolumn{2}{c}{Low: $\Delta\theta$, High: $0.1^\circ$} \\
        				& Source number estimators &\multicolumn{2}{c}{MDL, AIC} \\ 
       	% 				& \multirow{2}{*}{\shortstack[l]{Activation function\\hidden layers}} & \multicolumn{2}{l}{\multirow{2}{*}{ReLU}}  \\
    				% 	& & \multicolumn{2}{l}{ } \\
       	% 				& \multirow{2}{*}{\shortstack[l]{Activation function\\output layer}} & \multicolumn{2}{l}{\multirow{2}{*}{Softmax}}  \\
       	% 				& & \multicolumn{2}{l}{ } \\
        % 				& Loss function & \multicolumn{2}{l}{Sparse categorical cross-entropy}  \\
        % 				& Optimizer& \multicolumn{2}{l}{Adam}  \\ 
        % 				& Learning rate& \multicolumn{2}{l}{0.001}  \\ 
        % 				& Batch size & \multicolumn{2}{l}{32}\\ 
     			\bottomrule[1pt]
    													
    		\end{tabularx}
    	
    	\end{table}
    	
    	\subsubsection{Simulation Conditions}
    	
    	    The data for training and testing the proposed estimator are generated synthetically using the data model presented in Section~\ref{sec:data_model}. Two scenarios regarding the number of waves impinging at the sensor array are considered through numerical simulations: 
    	    \begin{description}
    	    \item[(I)] the source number $Q$ is assumed to be constant over all instances, i.e. $Q=2$, and
    	    \item[(II)] the source number $Q$ varies over the different instances, i.e., $Q$ is assumed to be a random variable drawn from the discrete uniform distribution $Q\sim U(1,4)$, meaning up to $4$ impinging waves are considered.
    	    \end{description}
	        The following have been assumed for both scenarios. A uniform linear array (ULA) of $N=8$ sensors with $\lambda/2$ inter-element spacing is considered, where $\lambda$ is the wavelength of the considered plane waves. The waves are uncorrelated and of equal power, i.e., $\mathbf{P}=\sigma^2 \mathbf{I}_Q$ \eqref{eq:sig_cov_mat}. The waves' AOAs are assumed to be random variables following the continuous uniform distribution, i.e., $\theta_1,\dots,\theta_Q\sim U(-60^\circ,60^\circ)$. The array's field of view (FOV) is defined by the interval $[\theta_\mathrm{min},\theta_\mathrm{max})=[-60^\circ, 60^\circ)$. The number of FOV segments evaluated by each classifier, i.e., the number of labels in a $k$-labelset, is set to $k=3$, as suggested for RA$k$EL\textsubscript{o} in \cite{tsoumakas2010random}. 
	        
	        For both scenarios, simulations are performed to investigate the impact of the signal-to-noise ratio (SNR) $\sigma^2 / \nu^2$, the FOV resolution (represented by $\Delta\theta$) and the number of layers in the framework, $L$. Specifically, the following values are considered: SNR $\in\{-10, 0, 10\}$ dB, $\Delta\theta\in\{2^{\circ},1^{\circ}\}$ (meaning $M=60$ and $M=120$, respectively \eqref{eq:del_theta}) and $L\in\{1,3,5\}$. 
    	   % The impact of the signal-to-noise ratio (SNR) $\sigma^2 / \nu^2$ is investigated for three specific values, i.e., -10, 0 and 10 dB. Furthermore, two FOV resolutions, $\Delta\theta\in\{2^{\circ},1^{\circ}\}$ , and three different values of $L$ (the number of layers in the framework), $L\in\{1,3,5\}$ are investigated to study their impact on the accuracy of the AOA estimates.
            % Since the array's FOV is discretised uniformly, each AOA could be within any of the $M$ grid intervals with equal probability.
            Hence, $2\times3\times2\times3=36$ (scenarios $\times$ SNRs $\times$ resolutions $\times$ framework layers) simulations are performed. Here, a 'simulation' comprises all three deployment stages presented in Section~\ref{subsec:train_test}. All random variables (source number $Q$, AOAs $\theta_1,\dots,\theta_Q$, waveforms $\mathbf{s}(t)$ and element noise $\mathbf{n}(t)$) follow the same distributions for all instances (an instance being a collection of $T$ snapshots of the array output) within a simulation, whether they are training or testing instances. New realizations are generated for each instance (source number and AOAs) and for each snapshot (waveforms and element noise) individually.
        
        \subsubsection{Learning-Parameters and Data Sets}
        
            In this work, the feedforward neural network (FFNN) (see, e.g., \cite{francois2017deep}) is employed as the single-label learning algorithm. The FFNN is one of the simplest type of neural networks (NNs) that exist, but still allows for sufficient design freedom to fit in the proposed AOA estimation framework. FFNNs are composed of an input layer, one or multiple hidden layers and an output layer. Each layer consists of a number of neurons. The number of neurons in the input layer is imposed by the dimension of the feature vectors. In the present work, each feature vector is composed as 
        	\begin{equation}
        		\label{eq:feature_vec}
        		\begin{split}
        			\mathbf{r}= [\hat{R}_{1,1}, \dots, \hat{R}_{N,N}, \Re(\hat{R}_{1,2}), \Im(\hat{R}_{1,2}),\\
        			\Re(\hat{R}_{1,3}), \Im(\hat{R}_{1,3}), \dots]^T,
        		\end{split}
        	\end{equation}
    	    where $\hat{R}_{i,j}$ is the element at row $i$ and column $j$ of $\mathbf{\hat{R}}$ \eqref{eq:Rhat}~\footnote{Since $\mathbf{\hat{R}}$ is Hermitian, only the diagonal elements and the elements on the upper right half of \eqref{eq:Rhat} are used. In fact, in case of isotropic sensors as considered here, the diagonal elements do not contain any information. Still, we include them in the feature vector, such that the impact of physically more realistic arrays can be easily investigated in the future.}. Hence, the number of neurons in the input layer equals $N^2$, with $N$ being the number of sensors in the array. Since the array data follow the Gaussian distribution, element-wise standardization is applied as the normalization algorithm, meaning all element-wise means and variances equal 0 and 1, respectively \cite{zheng2018feature}. The number of hidden layers and the number of neurons in these layers can be chosen freely. They are different for the different simulation scenarios, as can be seen in Table~\ref{tab:sim_params}, with the sequence of numbers representing the number of neurons in the hidden layers from input-side to output-side.
            All hidden layers are fully connected (i.e., each neuron is connected to all neurons in both the previous and the next layer) and use the ReLU activation function \cite{goodfellow2016deep}. The number of neurons in the output layer is imposed by the RA$k$EL parameter $k$ and equals $2^k$. The output layer uses the Softmax activation function (see, e.g., \cite{goodfellow2016deep}), meaning that all $2^k$ outputs are between 0 and 1 and add up to 1. Hence, they represent the probabilities $\tilde{P}_{j,\tilde{k}}$ ($j=1,\dots,m$ and $\tilde{k}=1,\dots,2^k$) which are converted to AOA estimates according to the procedure described in Section~\ref{subsec:estimates}.
            % In order to guarantee (at least theoretically) that the actual NN's outputs $\tilde{P}_{j,\tilde{k}}$ can achieve the target outputs $\tilde{S}_{j,\tilde{k}}$
	        % It should be noted that the number of antenna elements $N$ directly defines the number of features in the feature vectors and therefore also the number neurons in the input layer of each NN (i.e., $N^2$). In this work, it is assumed that the array layout is the same for all instances in the training set. Hence, the AOA estimator is specialised to one specific layout only. It can be seen from (\ref{eq:des_class_pred}) that the RA$k$EL parameter $k$ imposes the number of neurons in the output layer (namely $2^k$). 

            The NN training, i.e., the optimization of the NNs' weights, is performed using the Adam optimizer \cite{kingma2014adam} in combination with the categorical cross entropy loss function (see, e.g., \cite{francois2017deep}). The default \cite{kingma2014adam} learning rates of $\alpha=0.001$, $\beta_1=0.9$ and $\beta_2=0.999$ are used and each weight update is based on a mini-batch of 32 training instances. The training of a particular NN is terminated if the loss on the validation set did not decrease for 3 consecutive epochs (iterations over the training set). All simulations are implemented in Python using the TensorFlow machine learning library \cite{abadi2016tensorflow}.
            
            The evaluated threshold levels in the threshold optimization branch (Section~\ref{subsec:train_test}) are $0, 0.01, 0.02, \dots, 1$.
            
            Each instance, whether used for training or testing, comprises $T=100$ snapshots of the array output \eqref{eq:data_model}. The training set contains $D_{\mathrm{trn}}=80000$ instances for scenario I and $D_{\mathrm{trn}}=320000$ instance for scenario II. From all training instances, 80\% is used for training the classifiers, 10\% for validating them (i.e., determining when to stop training), and 10\% for optimizing the threshold level. In all simulations, the estimator is tested using $D_{\mathrm{tst}}=50000$ test instances. 

    	\subsubsection{Benchmark Algorithms}
            The joint AOA estimates obtained from the proposed estimator are compared (using the performance metrics presented in Section~\ref{sec:metrics}) to those obtained from the well-known MUSIC algorithm \cite{schmidt1986multiple}. Since the MUSIC algorithm belongs to the separable detection category, a source number estimate is required prior to estimating the AOAs. Two source number estimators are considered: the minimum description length (MDL) and the Akaike information criterion (AIC) \cite{wax1985detection}. For each simulation, the MUSIC angle spectrum is evaluated at two angle resolutions: (I) a lower resolution, equal to the FOV resolution $\Delta\theta$ of the proposed AOA estimation framework, and (II) a higher resolution of $0.1^\circ$. 

    \subsection{Results Simulation Scenario I: Fixed Source Number}
    \label{subsec:res_fixed}
        In this section, simulation results pertaining the fixed source number scenario are presented and analysed. 
            
        \subsubsection{Number of Framework Layers and FOV Resolution}
        \label{subsec:Q2_num_layers}
            \begin{figure}[tb]
    			\centering
    			\includegraphics[width=\columnwidth]{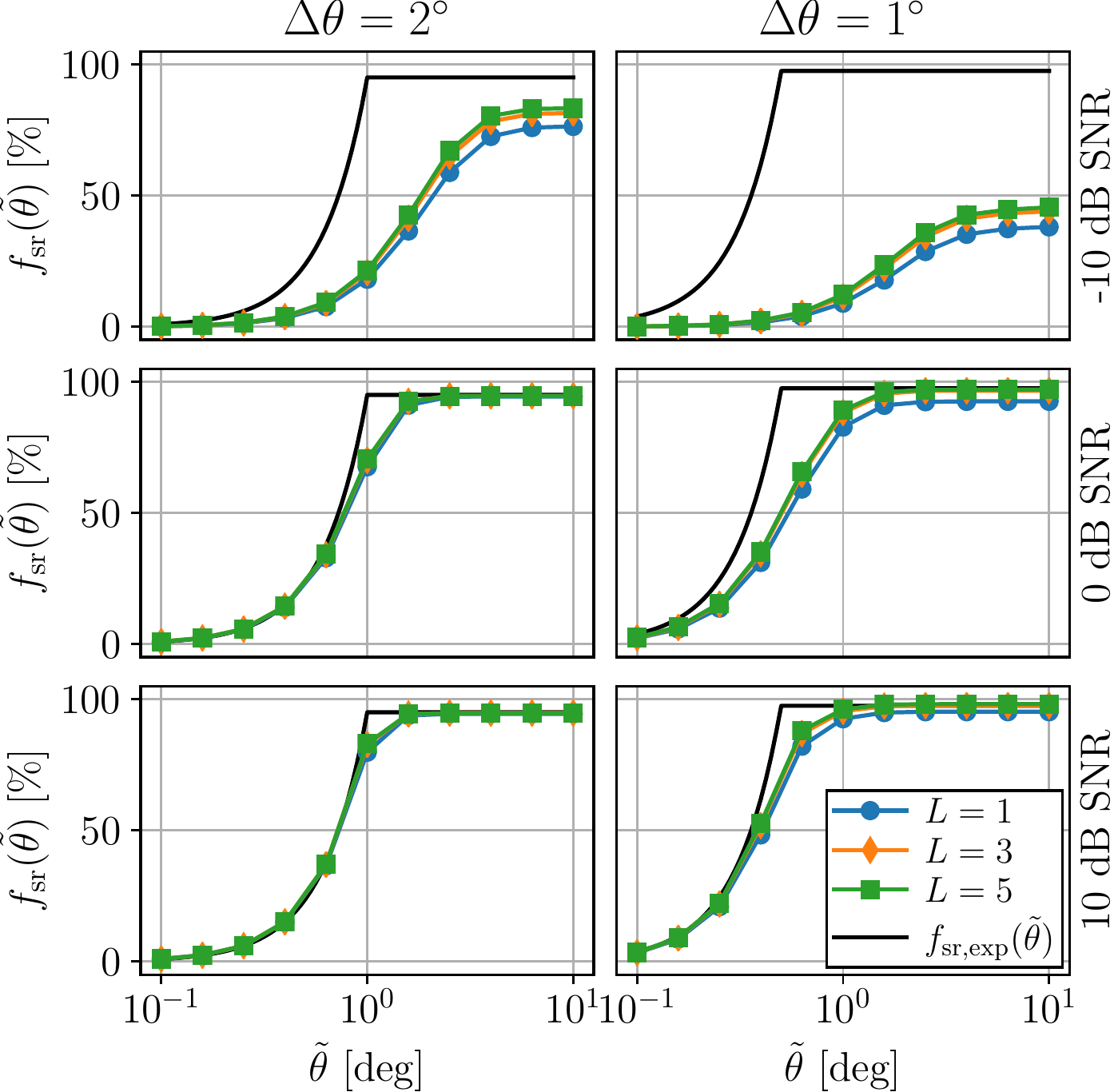}
    			\caption{Joint AOA estimation success rate $f_\mathrm{sr}(\tilde{\theta})$ of the proposed MLF vs. maximum allowed AOA estimation error $\tilde{\theta}$ for all simulations in scenario I. See Table~\ref{tab:sim_params}.}
    			\label{fig:Q2_num_layers}
    		\end{figure}
            Fig.~\ref{fig:Q2_num_layers} shows the joint AOA estimation success rate $f_\mathrm{sr}(\tilde{\theta})$ \eqref{eq:fsr} for various values of the maximum allowed AOA estimation error $\tilde{\theta}$, for all 18 simulations conducted within this scenario (3 SNRs $\times$ 2 values for $\Delta\theta$ $\times$ 3 values for $L$). The expected success rate in the case of ideal classifiers $f_{\mathrm{sr,exp}}(\tilde{\theta})$ \eqref{eq:success_rate}, which depends on $\Delta\theta$ but not on the SNR nor on $L$, is shown as a reference. As can be seen from Fig.~\ref{fig:Q2_num_layers}, increasing the number of framework layers $L$ increases the success rate $f_\mathrm{sr}(\tilde{\theta})$ for all the six considered $\{\mathrm{SNR},\Delta\theta\}$-couples, although the improvements are limited, especially when comparing $L=3$ and $L=5$. Hence, we conclude that the general recommendation of using RA$k$EL\textsubscript{o} with $M<m<2M$ and a small $k$ \cite{tsoumakas2010random}, equivalent to using $3<L<6$ for $k=3$ \eqref{eq:num_classifiers} in the layered framework proposed here, can be loosened for the present AOA application.

            % The number of classifiers in the framework increases linearly with $L$ (\ref{eq:num_classifiers}) and with that also the computational cost. 
            
            Fig.~\ref{fig:Q2_num_layers} also shows that for the two highest SNRs, the rate of successful AOA estimation is increased by using the higher FOV resolution ($\Delta\theta=1^\circ$) rather than the lower one ($\Delta\theta=2^\circ$), especially if $L\geq 3$.
            % all graphs (both the success rates and the references) shift to the upper-left when going from the lower to the higher resolution, i.e., from $\Delta\theta=2^\circ$ to $\Delta\theta=1^\circ$. In other words, for the two highest SNRs, the probability (or rate) of successful AOA estimation can be increased by using the higher resolution, where 'successful' implies that all AOA estimation errors are smaller than $\tilde{\theta}$. 
            For example, the success rate $\tilde{\theta}=1^\circ$ (i.e., assuming AOA estimation errors up to 1\textdegree~are allowed) increases from 70.6\% to 89.1\% (0 dB SNR, $L=5$) and from 83.1\% to 96.2\% (10 dB SNR, $L=5$). Considering the tightness of the reference $f_\mathrm{sr,exp}(\tilde{\theta})$ to the success rates achieved by the MLF at these SNRs for $\Delta\theta=2^\circ$ and the fact that the success rates increase (in absolute sense) when going to $\Delta\theta=1^\circ$, it is concluded that the performance of the MLF is limited by the FOV resolution when using $\Delta\theta=2^\circ$. On the contrary, when looking at the -10 dB SNR cases, it is observed that the success rate actually decreases when increasing the FOV resolution, e.g., from 21.4\% to 12.3\% for $L=5$ and $\tilde{\theta}=1^\circ$. It is worthwhile to note that the resolution increase is obtained at the expense of an increased computational cost. That is because the number of classifiers (here, NNs) to be trained is inversely proportional with $\Delta\theta$, see \eqref{eq:num_classifiers} and \eqref{eq:del_theta}. Hence, for the -10 dB SNR case, using the lower resolution is clearly the better option, both from the AOA estimation accuracy perspective as well as from the resource perspective.
            
            To get a better insight in the impact of the FOV resolution, we proceed by evaluating the predictive performance of the NNs by means of the averaged $F_1$-scores (Section~\ref{subsubsec:base_perf}).
            
        \subsubsection{Neural Network Predictive Performance}
        \label{subsec:Q2_NN_pred_perf}
            \begin{figure}[tbp]
               \centering
               \includegraphics[width=\columnwidth]{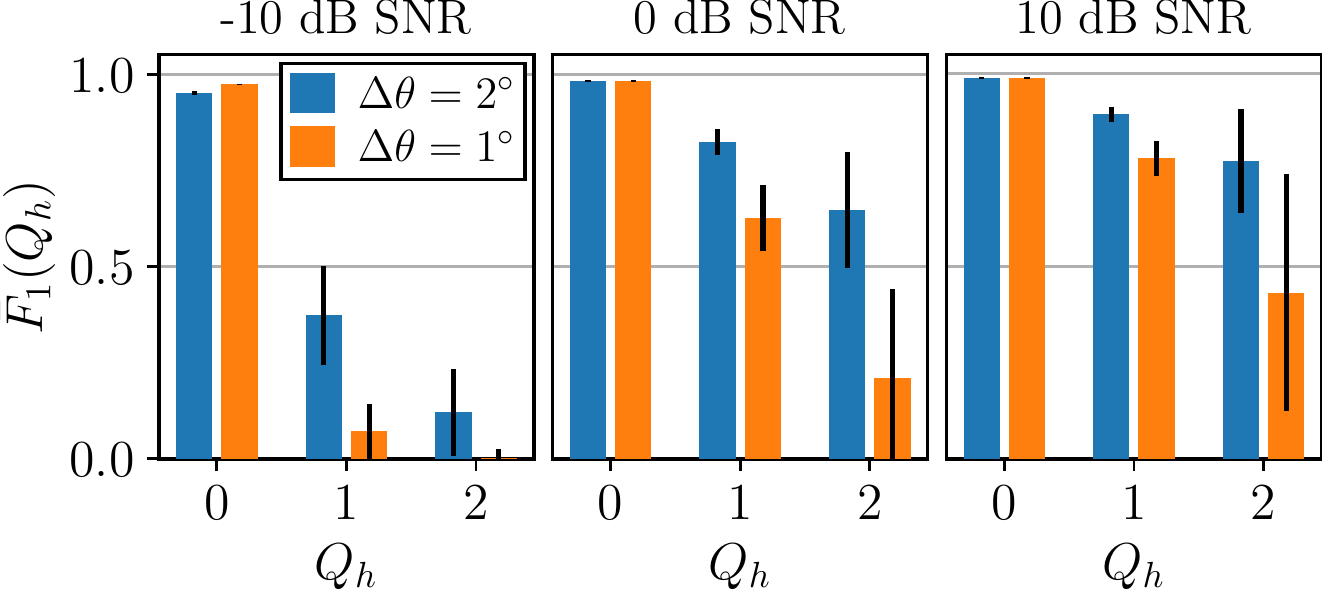}
                \caption{$F_1$-scores $\bar{F}_1(Q_h)$ of the proposed MLF vs. subset cardinality $Q_{h}$, for all simulations with $L=5$ in Scenario I. See Table~\ref{tab:sim_params}.}
                \label{fig:Q2_f1}
            \end{figure}
            Fig.~\ref{fig:Q2_f1} shows the $F_1$-scores $\bar{F}_1(Q_h)$ \eqref{eq:F1_bar} for the various subset cardinalities $Q_{h}=0,\dots,2$. Note that even though $k=3$, $\bar{F}_1(Q_{h}>2)$ is not defined because $Q=2$ in all simulations conducted within scenario I. The results presented in Fig.~\ref{fig:Q2_f1} are based on the simulations with $L=5$. Thus, $\bar{F}_1(Q_h)$ is computed by averaging the $F_1$-scores \eqref{eq:F1} of 100 and 200 NNs for the $\Delta\theta=2^\circ$ and $\Delta\theta=1^\circ$ frameworks, respectively \eqref{eq:num_classifiers}. As can be seen from Fig.~\ref{fig:Q2_f1}, $\bar{F}_1(Q_h)$ decreases when increasing the FOV resolution (i.e., decreasing $\Delta\theta$) for all SNRs and for all values of $Q_{h}$ except $Q_{h}=0$. This can be explained by a phenomenon called class imbalance \cite{galar2011review}. Although a detailed discussion is outside the scope of this work, it is worthwhile to note that this effect is expected to get stronger when further increasing the FOV resolution, as more and more instances from the training set will correspond to $Q_h=0$. Consequently, the NNs will have trouble learning an accurate mapping for instances corresponding to other values of $Q_h$. 
            % , which can be understood intuitively as follows. Given the simulation settings considered in this paper (i.e., a uniformly discretised FOV and uniformly distributed random AOAs), it is expected that for all classifiers (NNs), $((M-k)/M)^Q \times 100\% = ((M-3)/M)^2 \times 100\% $ of the training (and test) instances correspond to $Q_{h}=0$. Indeed, this percentage increases when increasing $M$, i.e., when increasing the FOV resolution by decreasing $\Delta\theta$. In other words, when increasing the resolution, the imbalance between the number of training instances representing the different values of $Q_h$ is increased. Learning an accurate mapping for the instances of $Q_h>0$ (the minorities) therefore becomes more and more complicated \cite{galar2011review}, resulting in a lower predictive performance for test instances corresponding to these values of $Q_h$. 
            
            While Fig.~\ref{fig:Q2_f1} shows that increasing the FOV resolution decreases the predictive performance at all considered SNRs, Fig.~\ref{fig:Q2_num_layers} shows that the joint AOA estimation success rate $f_{sr}(\tilde{\theta})$ only decreases at low SNR. This might sound paradoxical, but it is not: when increasing the FOV resolution while keeping the maximum allowed AOA estimation error $\tilde{\theta}$ fixed, one might (if $\tilde{\theta}>\Delta\theta/2$) obtain a successful AOA estimate also using non-perfect predictions. That this is indeed the case can be understood by evaluating the success rates relative to the references $f_\mathrm{sr,exp}(\tilde{\theta})$. As can be seen from Fig.~\ref{fig:Q2_num_layers} for $L=5$ and at mid and high SNR, the success rates are further apart from the references $f_\mathrm{sr,exp}(\tilde{\theta})$ for the higher FOV resolution ($\Delta\theta=1^\circ$) than for the lower FOV resolution ($\Delta\theta=2^\circ$). As the references $f_\mathrm{sr,exp}(\tilde{\theta})$ assume ideal classifiers, this indicates that indeed the NNs' predictions are further from ideal for the higher FOV resolution, as confirmed by results shown in Fig.~\ref{fig:Q2_f1}.
            % This can be seen by comparing the success rates to the references $f_\mathrm{sr,exp}(\tilde{\theta})$ (which assume ideal classifiers) at mid and high SNR and at both FOV resolutions (Fig.~\ref{fig:Q2_num_layers}). Clearly, the success rates are further apart from the reference at the higher FOV resolution, which indicates that indeed the NNs' predictions are further from ideal as confirmed by results shown in Fig.~\ref{fig:Q2_f1}. However, this does not mean that the success rate cannot increase in the absolute sense.  
            Interestingly, it is found that the threshold level, which is optimized during the training stage, increased from 0.05 to 0.22 (0 dB SNR) and from 0.04 to 0.23 (10 dB SNR) when increasing the FOV resolution from $\Delta\theta=2^\circ$ to $\Delta\theta=1^\circ$. This indicates that at the higher resolution, there are peaks in the probabilistic angle spectra at angles other than the AOAs that need to be filtered out. This is a direct consequence of incorrect classifier predictions. At low SNR (-10 dB), the situation is different, as the decreased predictive performance resulting from an increased FOV resolution caused the success rate $f_{sr}(\tilde{\theta})$ to decrease in absolute sense as well. More simulations are required to investigate if this also occurs when further increasing the FOV resolution at mid and high SNR, and, if so, to find the optimum FOV resolution for a given SNR.
            
            % Since the threshold is optimized to maximize the number of instances for which the source number estimate is correct, this indicates that for the $\Delta\theta=1^\circ$ framework, there are additional spectrum peaks at unwanted angles (originating from the incorrect classifier predictions) that need to be filtered out. Hence, the threshold can compensate for the reduced predictive performance, but only if the peaks at the desired angles are higher than those unwanted peaks. The latter is clearly not the case in the -10 dB SNR case. 
            
            % At the same time though, it was shown (Fig.~\ref{fig:Q2_num_layers}) that the \{-10 dB SNR, $\Delta\theta=1^\circ$\}-couple is the only one in which the success rate clearly increases by using larger values of $L$. In other words, the negative impact of the increased framework resolution (decreased $\Delta \theta$) on the success rate can be compensated for, to some extent, by increasing the number of framework layers $L$. However, as both increasing the resolution as well as increasing $L$ is at the expense of an increased computational load, using the lower resolution is definitely considered the better option for the -10 dB SNR case. More extensive simulations are required to determine the optimal resolution (yielding maximum success rate with minimal computational cost) for a given SNR.

    	\subsubsection{Benchmark Comparison}
    	\label{subsubsec:Q2_benchmarks}
    	
    		\setlength{\tabcolsep}{4pt}
    		\begin{table}[tbp]
    			\centering	
    			\caption{$P(\hat{Q}=Q)$ and RMSE for Scenario I. See Table~\ref{tab:sim_params}.}
    			\label{tab:Q2_PQQhat}
    			\begin{tabularx}{\columnwidth}{@{}lX@{}cccccc@{}}
    				\toprule[1pt]
    				                                                    & SNR [dB]              & \multicolumn{2}{c}{-10}   & \multicolumn{2}{c}{0}     & \multicolumn{2}{c}{10}\\ \cmidrule{2-8}
    				                                                    & $\Delta\theta$ [deg] 	& 2 	            & 1 	& 2		& 1 	            & 2 	& 1 \\ \midrule 		
    				\multirow{3}{5em}{\centering $P(\hat{Q}=Q)$ $[\%]$} & MLF, $L=5$ 			& \textbf{83.5} 	& 50.8	& 94.5 	& \textbf{97.1} 	& 94.7	& \textbf{98.1} \\ 
    				                                                    & MDL 					& \multicolumn{2}{c}{0.5}   &\multicolumn{2}{c}{91.9}   &\multicolumn{2}{c}{97.4} \\
    				                                                    & AIC 					& \multicolumn{2}{c}{63.7}  &\multicolumn{2}{c}{86.8}   & \multicolumn{2}{c}{89.8} \\ \midrule 
    				
    				\multirow{5}{5em}{\centering RMSE [deg]}            & MLF, $L=5$ 			        & \textbf{2.2} & 13.1        & 0.7  & \textbf{0.5}      & 0.6 & \textbf{0.4}\\ 
    				                                                    & MUSIC ($\Delta\theta$) + MDL 	& 3.7 & 3.6                 & 7.5  & 7.2                & 8.5 & 8.1\\
                                                        				& MUSIC ($0.1^\circ$) + MDL 	& \multicolumn{2}{c}{3.6}   & \multicolumn{2}{c}{7.2}   & \multicolumn{2}{c}{8.0}\\
                                                        				& MUSIC ($\Delta\theta$) + AIC 	& 9.5 & 9.4                 & 10.3 & 10.2               & 9.4 & 9.1\\
                                                        				& MUSIC ($0.1^\circ$) + AIC 	& \multicolumn{2}{c}{9.4}   & \multicolumn{2}{c}{10.1}  & \multicolumn{2}{c}{9.1}\\
    				\bottomrule[1pt]
    			\end{tabularx}
    		\end{table} 
    		
            In this section, the joint AOA estimates of the proposed MLF are compared to those attained from the reference algorithms MDL, AIC (source number estimates) and MUSIC (AOA estimates). Again, the results presented for the MLF are based on the $L=5$ simulations.
            
            Table~\ref{tab:Q2_PQQhat} presents the source number estimation accuracy $P(\hat{Q}=Q)$ and the root-mean-square error (RMSE) for all considered $\{\mathrm{SNR},\Delta\theta\}$-couples. The best performing algorithm (i.e., the one achieving the highest $P(\hat{Q}=Q)$ and the lowest RMSE) is highlighted in bold for each SNR. In terms of both metrics, the MLF outperforms the benchmark algorithms for all considered SNRs, although this requires different FOV resolutions: It is observed once more that the low resolution ($\Delta\theta=2^\circ$) is preferred for the -10 dB SNR case, whereas the high resolution ($\Delta\theta=1^\circ$) achieves better results at the mid and high range SNRs. 
    		
    		\begin{figure}[tbp]
    			\centering
    			\includegraphics[width=\columnwidth]{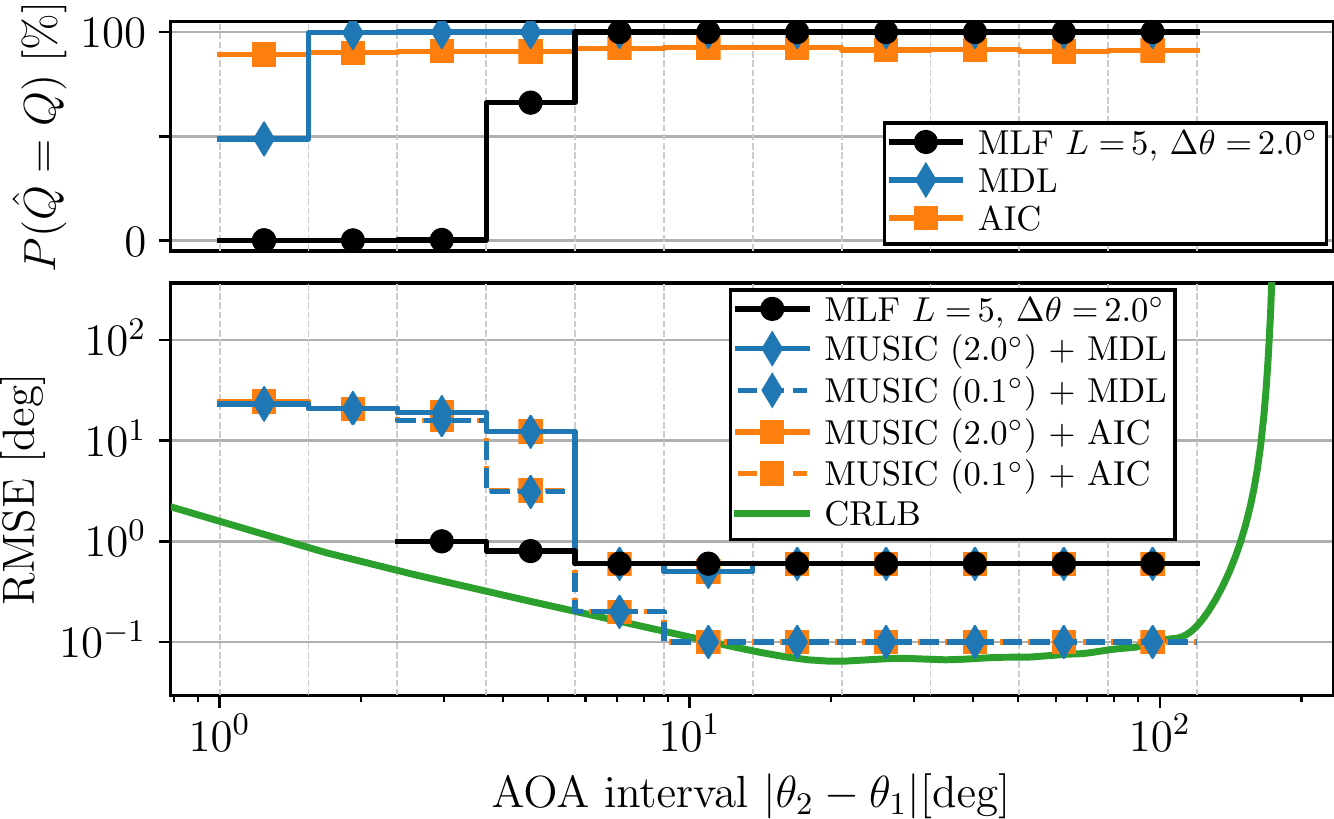}
    			\caption{$P(\hat{Q}=Q)$ and RMSE vs. AOA interval. 10 dB SNR, $\Delta\theta=2^\circ$ and $L=5$, Scenario I. Results are based on a test set in which the sources' AOAs are symmetric with respect to broadside. All other parameters are as in Table~\ref{tab:sim_params}, Scenario I.}
    			\label{fig:Q2_RMSE_vs_int}
    		\end{figure}
    		To clarify the relatively high RMSEs for the MUSIC algorithm, we plot both $P(\hat{Q}=Q)$ and the RMSE against the AOA interval $|\theta_2 - \theta_1|$ in Fig.~\ref{fig:Q2_RMSE_vs_int}. To this end, we synthesized additional test sets (12000 instances) in which the AOAs of the two sources are symmetric with respect to the array's broadside, i.e., $\theta_q=90 \pm \delta$ degree. All other parameters are as in Table~\ref{tab:sim_params}. The AOA interval $|\theta_2 - \theta_1|=|90+\delta - (90-\delta)|=2\delta$ is assumed to be a random variable following a continuous log-uniform probability distribution between 1\textdegree~and 120\textdegree. We grouped the instances in these test sets based on their AOA interval and computed $P(\hat{Q}=Q)$ and the RMSE for each group separately, as indicated by the vertical grid and the stair-wise graphs in Fig.~\ref{fig:Q2_RMSE_vs_int}. In this way, we 'average out' (especially at large AOA intervals) the impact of the finite resolution which is inherent to both the MLF and the MUSIC algorithm. As an additional reference, the Cramér-Rao lower bound (CRLB), see, e.g., \cite{van2004optimum,stoica1989music}, is shown as well. For the sake of conciseness, we only present results for the \{10 dB SNR, $\Delta\theta=2^\circ$\}-couple, but similar observations were made in the other considered cases as well. 
    % 		we plot both $P(\hat{Q}=Q)$ and the RMSE against the AOA interval $|\theta_2 - \theta_1|$ in Fig.~\ref{fig:Q2_RMSE_vs_int}. To this end, we grouped the instances in the test set based on their AOA interval, whereupon $P(\hat{Q}=Q)$ and the RMSE are computed for each group separately. Hence, the AOA interval indicated by any data point in Fig.~\ref{fig:Q2_RMSE_vs_int} represents the average AOA interval of all instances within a single group. Only AOA intervals larger than 1\textdegree~are considered, as smaller intervals have a small support when grouping based on a log-scale because of the uniformly distributed random AOAs. For the sake of conciseness, we only show results for the \{10 dB SNR, $\Delta\theta=2^\circ$\}-couple, but similar observations were made in the other considered cases as well. 
    		As can be seen from Fig.~\ref{fig:Q2_RMSE_vs_int}, MDL and AIC outperform the MLF at small AOA intervals. This is because in this specific symmetric scenario, an AOA interval of at least $2\Delta\theta=4^\circ$ is required for the MLF to be able to resolve both sources (Section~\ref{subsubsec:estimation}). Hence, at these small AOA intervals, the MLF never estimates the source number correctly and therefore, the RMSE cannot be computed. Contrarily, the RMSE for the MUSIC algorithm \textit{does} exist at small intervals, although it is nearly 2 orders of magnitude larger than the CRLB (worst case). This can be understood as follows. Since the MUSIC algorithm belongs to the separable detection category, it aims to return as many AOA estimates as required according to the source number detection method, here MDL/AIC. In case MDL/AIC manages to estimate the correct source number, while at the same time the MUSIC angle spectrum does not contain distinct peaks at all AOAs (which might happen for small AOA intervals \cite{krim1996two}), the argument of another peak in the spectrum is returned. This results in large AOA estimation errors, which dominate the RMSE values presented in Table~\ref{tab:Q2_PQQhat}. This phenomenon emphasizes the advantage of the proposed joint AOA estimation success rate $f_\mathrm{sr}(\tilde{\theta})$ \eqref{eq:fsr}, as this metric considers both source number and the AOA estimates. Hence, next we compare the proposed MLF and the MUSIC algorithm (combined with MDL/AIC) in terms of the joint AOA estimation success rate.
            % In other words, if the MLF correctly estimates the source number, the estimated AOAs are accurate as well. Whether they are accurate enough depends on the maximum allowed error $\tilde{\theta}$ that can be defined by the user. 
            
    		\begin{figure}[tbp]
    			\centering
    			\includegraphics[width=\columnwidth]{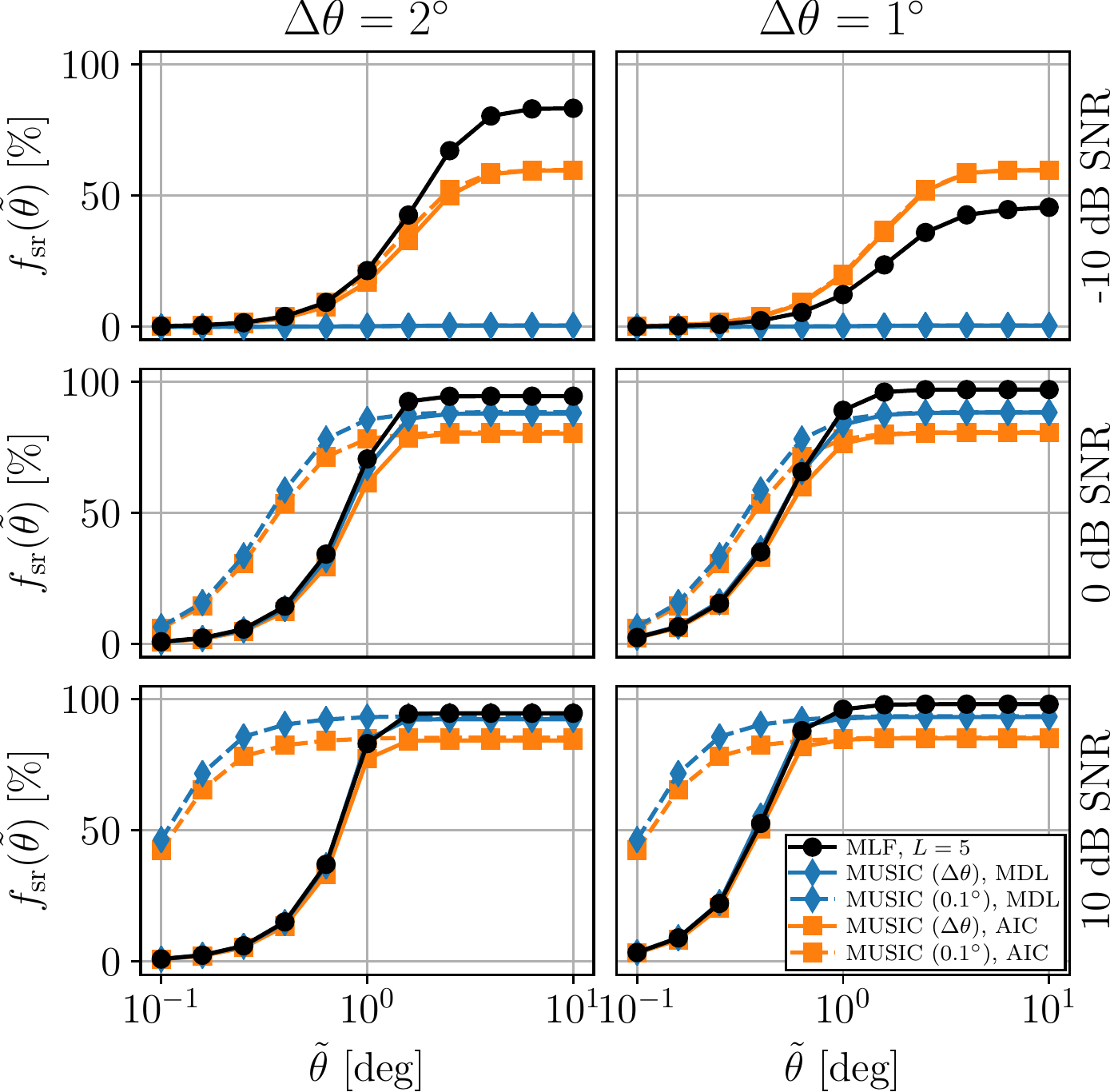}
    			\caption{Joint AOA estimation success rate $f_\mathrm{sr}(\tilde{\theta})$ vs. maximum allowed AOA estimation error $\tilde{\theta}$, proposed MLF (with $L=5$) vs. MUSIC for simulations in Scenario I. See Table~\ref{tab:sim_params}.}
    			\label{fig:Q2_SR_MUSIC}
    		\end{figure}
    		
            As can be seen in in Fig.~\ref{fig:Q2_SR_MUSIC}, the proposed MLF outperforms the MUSIC algorithm if $\tilde{\theta} \gtrapprox \Delta\theta$, i.e., if the maximum allowed AOA estimation error is approximately of the same order as (or larger than) the size of the FOV segments. This applies to all variants of the MUSIC algorithm considered (low/high angle spectrum resolution, see Table~\ref{tab:sim_params}, and MDL/AIC source number detection) and to almost all $\{\mathrm{SNR},\Delta\theta\}$-couples. Only for the \{-10 dB SNR, $\Delta\theta=1^\circ$\}-couple, the MUSIC+AIC combination attains a higher success rate than the MLF. Contrarily, if $\tilde{\theta}<\Delta\theta$, the high resolution MUSIC algorithm outperforms the MLF for the mid and high SNRs. This is a direct consequence of the finite FOV resolution of the MLF, because of which errors up to $\Delta\theta/2$ are to be expected, as already illustrated by $f_\mathrm{sr,exp}(\tilde{\theta})$ \eqref{eq:success_rate} in Fig.~\ref{fig:Q2_num_layers}. Additional simulations are required to determine whether a framework with a higher FOV resolution can outperform the $0.1^\circ$ MUSIC algorithm also for small $\tilde{\theta}$. 
            
            % \textcolor{red}{To get a fair view on the impact of NNs' predictive performance on the success rate achieved by frameworks with different resolutions, one would have to evaluate the success rates at the same \textit{relative} maximum allowed error. However, as the classifiers are just a means to obtain the AOA estimates (i.e., achieving a high predictive performance is not the main goal itself), it was decided to show the success rate for the same \textit{absolute} maximum allowed error.}
    		
    \subsection{Results Simulation Scenario II: Variable Source Number}
    \label{subsec:res_source_number}
        Next, we present an analysis of the simulation results pertaining the variable source number scenario. For the sake of conciseness, we limit ourselves to the benchmark comparison, as the phenomena observed in Section~\ref{subsec:res_fixed}, e.g., limited improvements for $L>3$ and a decreasing predictive performance for increasing $Q_h$ due to class imbalance, apply here as well. 
        
        \subsubsection{Benchmark Comparison}

		\begin{figure}[tbp]
			\centering
			\includegraphics[width=\columnwidth]{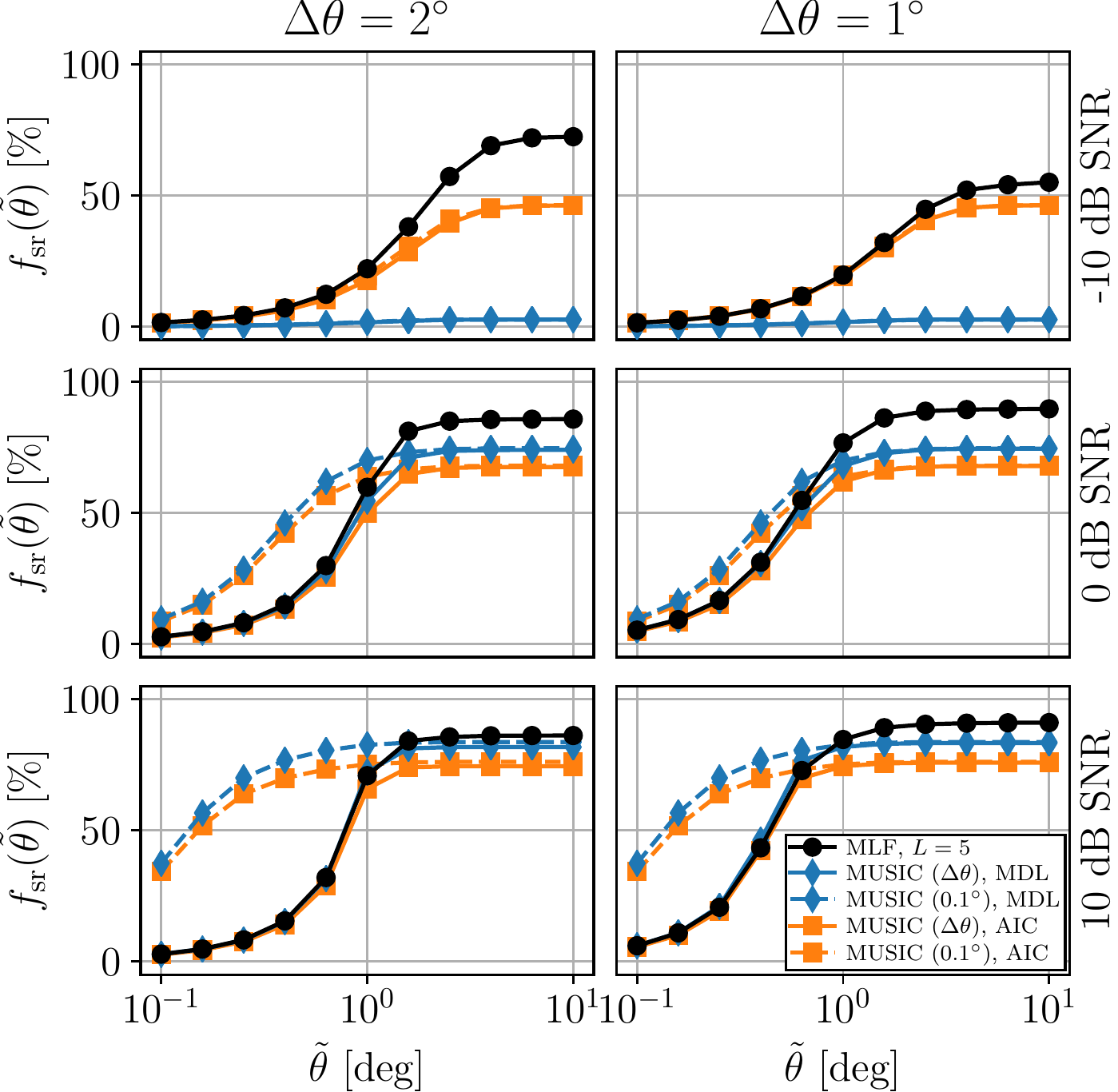}
			\caption{Joint AOA estimation success rate $f_\mathrm{sr}(\tilde{\theta})$ vs. maximum allowed AOA estimation error $\tilde{\theta}$, proposed MLF (with $L=5$) vs. MUSIC for simulations in Scenario II. See Table~\ref{tab:sim_params}.}
			\label{fig:Q14_SR_MUSIC}
		\end{figure}
	    
        Fig.~\ref{fig:Q14_SR_MUSIC} shows the joint AOA estimation success rate $f_\mathrm{sr}(\tilde{\theta})$, plotted against maximum allowed AOA estimation error $\tilde{\theta}$, for all considered $\{\mathrm{SNR},\Delta\theta\}$-couples. Again, the results shown for the MLF were obtained using a framework with $L=5$ layers. As can be seen, the success rates for the MLF and for the MUSIC algorithm follow the same trends as in scenario I (Fig.~\ref{fig:Q2_SR_MUSIC}), although they have decreased in absolute sense for all values of $\tilde{\theta}$ for both algorithms. Contrary to scenario I, the MLF now outperforms the MUSIC-AIC combination in the \{-10 dB SNR, $\Delta\theta=1^\circ$\}-case as well. Nevertheless, still the $\Delta\theta=2^\circ$ MLF achieves higher success rates than the $\Delta\theta=1^\circ$ MLF at this low SNR. 
		
    	\begin{figure}[tbp]
    		\centering
    		\includegraphics[width=\columnwidth]{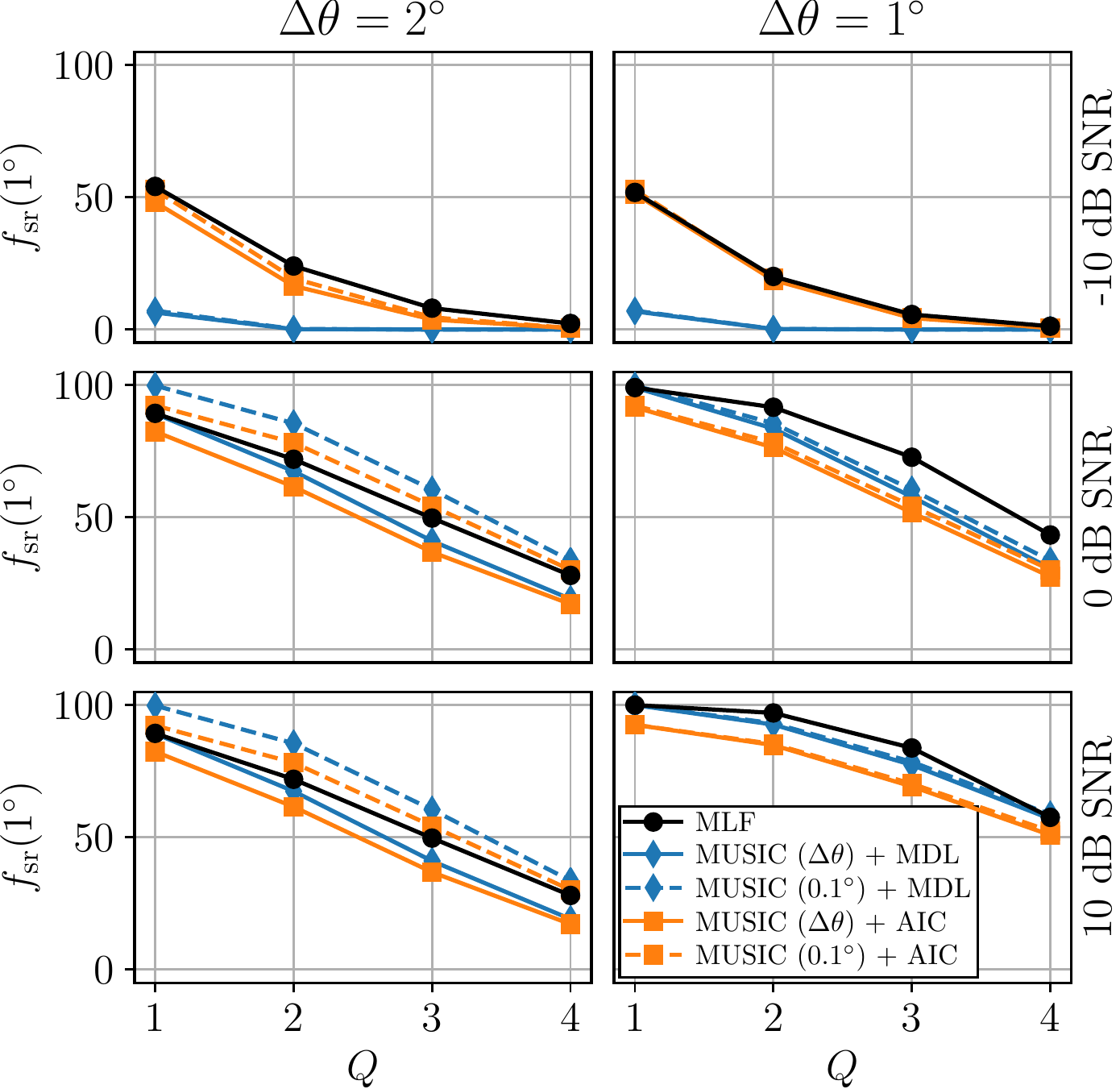}
    		\caption{Success rate $f_\mathrm{sr}(1^\circ)$ vs. source number $Q$ for MLF and MUSIC. Scenario II.}
    		\label{fig:Q14_SR_vs_Q}
    	\end{figure}
		
		In order to get a better insight into the impact of the various source numbers on the estimator performance, we group all test instances based on the number of sources $Q$ and evaluate the success rate $f_\mathrm{sr}(\tilde{\theta}=1^\circ)$ for each of them separately (Fig.~\ref{fig:Q14_SR_vs_Q}). Hence, AOA estimation errors up to $1^\circ$ are considered acceptable. It is worthwhile to note that the relative maximum allowed AOA estimation error $\tilde{\theta}/\Delta\theta$ is larger for the high resolution framework ($\Delta\theta=1^\circ$) than for the low resolution framework ($\Delta\theta=2^\circ$). Consequently, a higher success rate can be achieved by the high resolution MLF, even though the NNs have a lower predictive performance (not shown for the sake of conciseness) than those in the low resolution MLF. This was also observed in scenario I, for SNRs equal to 0 and 10 dB (see Fig.~\ref{fig:Q2_SR_MUSIC} at $\tilde{\theta}=1^\circ$ and Fig.~\ref{fig:Q2_f1}). As can be seen from Fig.~\ref{fig:Q14_SR_vs_Q}, the success rate decreases for increasing source numbers $Q$, both for the MLF and the MUSIC algorithm. We conclude that in this simulation scenario and for this particular maximum allowed AOA estimation error ($\tilde{\theta}=1^\circ$), only the MLF with low FOV resolution ($\Delta\theta=2^\circ$) is outperformed by the (high resolution) MUSIC algorithm, and only at SNRs of 0 and 10 dB. In all other cases, the MLF achieves the highest success rate for all considered values of $Q$.

\section{Conclusion}
\label{sec:conclusions}
    In this paper, we proposed a machine learning framework (MLF) which jointly estimates the source number and the angles-of-arrival (AOAs) of plane waves impinging a sensor array. The MLF is tailored to the array's segmented field of view (FOV) such that it can solve the joint AOA estimation problem through supervised-learning-based classification. The proposed approach is general in the sense that the MLF can, in principle, be implemented in combination with any single-label multi-class classification algorithm. Moreover, a new performance metric, the joint AOA estimation success rate, is introduced to assess the performance of the proposed MLF. Particularly, this metric depends on the user-defined maximum allowed AOA estimation error.
    % Here, we specialized our analysis to feedforward neural networks because they are straightforward to implement and they have been studied thoroughly. The proposed estimator is evaluated through numerical simulations. Two different simulation scenarios representing (I) a fixed source number and (II) a variable source number have been studied. 
    Numerical simulations are conducted using feedforward neural networks as the learning algorithm. In scenarios representing both fixed and variable source numbers, results show that the joint AOA estimation success rate attained by the MLF strongly depends on the resolution of the FOV segmentation (the FOV resolution). When increasing the FOV resolution from 2\textdegree~to 1\textdegree~while keeping the learning settings the same, the achieved success rate deteriorates at low signal-to-noise ratio (SNR) of -10 dB, for all considered values of the maximum allowed AOA estimation error. On the contrary, at mid (0 dB) and high (10 dB) SNRs, the success rate increases when increasing the FOV resolution. The FOV resolution is inversely proportional to the number of classifiers in the MLF. Hence, an important trade-off between the estimation performance and the computational burden is to be considered, especially at mid and high range SNRs. In nearly all considered cases, the MLF outperforms the multiple signal classification (MUSIC) algorithm, implemented in conjunction with the source number estimator Akaike’s information criterion (AIC) or the minimum description length (MDL). Only in case of a fixed source number and at low SNR, the MLF with high FOV resolution (1\textdegree) is outperformed by the MUSIC-AIC combination. We conclude that the proposed MLF offers a higher rate of successful joint AOA estimation for all SNRs if the maximum allowed AOA estimation error is of the order of (or larger than) the size of the FOV segments and if the FOV resolution is selected with care.

    % Results show that the MLF outperforms the reference algorithm in terms of the success rate for almost all considered cases (especially those with a variable source number), as long as the maximum allowed AOA estimation error is about equal to (or larger than) the size of the FOV segments. The latter does only not apply for the MLF with the higher FOV resolution at -10 dB SNR in scenario (I), where the MUSIC algorithm combined with AIC achieves higher success rates. Hence, we emphasize once more that, at this low SNR, the lower FOV resolution is preferred. 
    
    Further research into the relation between different design parameters, e.g., the FOV resolution, the number of framework layers, the learning algorithm and the number of training instances, and the properties of the signal environment, e.g., source number and SNR, are required to determine the optimal framework topology for a given scenario. Moreover, further investigation into the use of different FOV discretisations for the different framework layers is recommended. Finally, the study of the impact of the physical properties of realistic sensor arrays on the estimation accuracy of the proposed estimator in comparison to conventional estimators like the MUSIC algorithm is of great interest for practical applications.
\appendices

\section{Expected Success Rate Ideal Classifiers}
\label{sec:App_A}
    
    Consider the independent and identically distributed random variables $\theta_1,\dots,\theta_Q$, drawn from the continuous uniform distribution $U(\theta_\mathrm{min},\theta_\mathrm{max})$. Assume the interval $[\theta_\mathrm{min}, \theta_\mathrm{max})$ is segmented in $M$ intervals $[\theta_\mathrm{i,min}, \theta_\mathrm{i,max})$ ($i=1,\dots,M$) in a regular manner, meaning each interval has size $\Delta\theta=(\theta_\mathrm{max} - \theta_\mathrm{min})/M$. Let's denote the center of the $i$\textsuperscript{th} interval $c_i=(\theta_\mathrm{i,min}+\theta_\mathrm{i,max})/2$. Then, the probability $P$ that all $\theta_1,\dots,\theta_Q$ are at most $\tilde{\theta}$ removed from one of the interval centers $c_1,\dots,c_M$ (for $\tilde{\theta}<\Delta\theta/2$) is computed as
	\begin{equation}
		\begin{split}
			P(\tilde{\theta})\big\rvert_{\tilde{\theta}<\Delta\theta/2} &= \prod_{q=1}^{Q} M \int_{-\tilde{\theta}}^{\tilde{\theta}} \frac{1}{\theta_\mathrm{max}-\theta_\mathrm{min}} d\theta_q \\
			  &=\bigg[ \frac{\theta_\mathrm{max}-\theta_\mathrm{min}}{\Delta\theta}\int_{-\tilde{\theta}}^{\tilde{\theta}} \frac{1}{\theta_\mathrm{max}-\theta_\mathrm{min}} d\theta \bigg]^Q \\
			  &= \bigg[\frac{\tilde{\theta}}{\Delta\theta/2}\bigg]^Q.
		\end{split}
	\end{equation}
	Clearly, if $\tilde{\theta}\geq \Delta\theta/2$, $P({\tilde{\theta}})=1$, since the closest $c_i$ is at a distance of at most $\Delta\theta/2$ from any point in the interval $[\theta_\mathrm{min}, \theta_\mathrm{max})$. Hence, it follows that
	\begin{equation}
           P(\tilde{\theta})=
            \begin{cases}
            \Big(\frac{\tilde{\theta}}{\Delta\theta/2}\Big)^Q & \text{if } \tilde{\theta} < \Delta\theta/2 \\
             1 & \text{otherwise}.
            \end{cases}
        \end{equation}
	
% 	The expected success rate for the estimator with a maximum allowed AOA estimation error of  $\tilde{\theta}<\Delta\theta/2$ therefore equals $\big(\frac{\tilde{\theta}}{\Delta\theta/2}\big)^Q f_{\mathrm{sr,exp},\mathrm{max}}$ 
	
% 	Now consider that $\tilde{\theta} < \Delta\theta/2$. In this case, an additional requirement needs to be fulfilled for the estimator to be successful, being that all $Q$ AOAs need to be sufficiently close to one of the segment centres. This is only a matter of chance. In case of uniformly distributed random AOAs and a uniform grid, both sharing the same interval, this probability equals

\section{Maximum Expected Success Rate}
\label{sec:App_B}

    Consider performing random sampling with replacement from the set $\{\lambda_1,\dots,\lambda_M\}$, where the likelihood of selecting a particular $\lambda_i$ ($i=1,\dots,M$) is equal for all of them. Hence, when sampling $Q$ times, $M^Q$ possible outcomes (permutations) exist. Assume we want to compute the percentage $p$ of these $M^Q$ permutations which fulfill the requirements that (I) none of the $\lambda_i$ is selected multiple times, and (II) no neighbouring $\lambda_i$ are selected, i.e., when $\lambda_i$ is selected, $\lambda_{i-1}$ and $\lambda_{i+1}$ are not. Here, the latter requirement reduces to either $\lambda_{i+1}$ or $\lambda_{i-1}$ if $i=1$ or $i=M$, respectively. This can be interpreted as random sampling without replacement $Q$ times from a set of $M-(Q-1)$ elements, for which the number of combinations equals ${M - (Q-1) \choose Q}$. Multiplying this by $Q!$ converts the combinations to permutations, meaning that the percentage of permutations fulfilling requirements (I) and (II) is computed as
	\begin{equation}
		p = {M - (Q-1) \choose Q} \frac{Q!}{M^Q} \times 100\%.
	\end{equation}

\ifCLASSOPTIONcaptionsoff
  \newpage
\fi

% \bibliographystyle{IEEEtran}
% \printbibliography

\bibliographystyle{IEEEtran}
\bibliography{IEEEabrv,references}

\begin{thebibliography}{10}
\providecommand{\url}[1]{#1}
\csname url@samestyle\endcsname
\providecommand{\newblock}{\relax}
\providecommand{\bibinfo}[2]{#2}
\providecommand{\BIBentrySTDinterwordspacing}{\spaceskip=0pt\relax}
\providecommand{\BIBentryALTinterwordstretchfactor}{4}
\providecommand{\BIBentryALTinterwordspacing}{\spaceskip=\fontdimen2\font plus
\BIBentryALTinterwordstretchfactor\fontdimen3\font minus
  \fontdimen4\font\relax}
\providecommand{\BIBforeignlanguage}[2]{{%
\expandafter\ifx\csname l@#1\endcsname\relax
\typeout{** WARNING: IEEEtran.bst: No hyphenation pattern has been}%
\typeout{** loaded for the language `#1'. Using the pattern for}%
\typeout{** the default language instead.}%
\else
\language=\csname l@#1\endcsname
\fi
#2}}
\providecommand{\BIBdecl}{\relax}
\BIBdecl

\bibitem{krim1996two}
H.~Krim and M.~Viberg, ``Two decades of array signal processing research: the
  parametric approach,'' \emph{IEEE signal processing magazine}, vol.~13,
  no.~4, pp. 67--94, 1996.

\bibitem{van2004optimum}
H.~L. Van~Trees, \emph{Optimum array processing: Part IV of detection,
  estimation, and modulation theory}.\hskip 1em plus 0.5em minus 0.4em\relax
  John Wiley \& Sons, 2004.

\bibitem{wax1985detection}
M.~Wax and T.~Kailath, ``Detection of signals by information theoretic
  criteria,'' \emph{IEEE Transactions on acoustics, speech, and signal
  processing}, vol.~33, no.~2, pp. 387--392, 1985.

\bibitem{capon1969high}
J.~Capon, ``High-resolution frequency-wavenumber spectrum analysis,''
  \emph{Proceedings of the IEEE}, vol.~57, no.~8, pp. 1408--1418, 1969.

\bibitem{schmidt1986multiple}
R.~Schmidt, ``Multiple emitter location and signal parameter estimation,''
  \emph{IEEE transactions on antennas and propagation}, vol.~34, no.~3, pp.
  276--280, 1986.

\bibitem{roy1989esprit}
R.~Roy and T.~Kailath, ``Esprit-estimation of signal parameters via rotational
  invariance techniques,'' \emph{IEEE Transactions on acoustics, speech, and
  signal processing}, vol.~37, no.~7, pp. 984--995, 1989.

\bibitem{barabell1983improving}
A.~Barabell, ``Improving the resolution performance of eigenstructure-based
  direction-finding algorithms,'' in \emph{ICASSP'83. IEEE International
  Conference on Acoustics, Speech, and Signal Processing}, vol.~8.\hskip 1em
  plus 0.5em minus 0.4em\relax IEEE, 1983, pp. 336--339.

\bibitem{stoica1989music}
P.~Stoica and A.~Nehorai, ``Music, maximum likelihood, and cramer-rao bound,''
  \emph{IEEE Transactions on Acoustics, speech, and signal processing},
  vol.~37, no.~5, pp. 720--741, 1989.

\bibitem{ziskind1988maximum}
I.~Ziskind and M.~Wax, ``Maximum likelihood localization of multiple sources by
  alternating projection,'' \emph{IEEE Transactions on Acoustics, Speech, and
  Signal Processing}, vol.~36, no.~10, pp. 1553--1560, 1988.

\bibitem{yang2012off}
Z.~Yang, L.~Xie, and C.~Zhang, ``Off-grid direction of arrival estimation using
  sparse bayesian inference,'' \emph{IEEE Transactions on Signal Processing},
  vol.~61, no.~1, pp. 38--43, 2012.

\bibitem{chen2018off}
P.~Chen, Z.~Cao, Z.~Chen, and X.~Wang, ``Off-grid doa estimation using sparse
  bayesian learning in mimo radar with unknown mutual coupling,'' \emph{IEEE
  Transactions on Signal Processing}, vol.~67, no.~1, pp. 208--220, 2018.

\bibitem{liu2012efficient}
Z.-M. Liu, Z.-T. Huang, and Y.-Y. Zhou, ``An efficient maximum likelihood
  method for direction-of-arrival estimation via sparse bayesian learning,''
  \emph{IEEE Transactions on Wireless Communications}, vol.~11, no.~10, pp.
  1--11, 2012.

\bibitem{yang2018sparse}
Z.~Yang, J.~Li, P.~Stoica, and L.~Xie, ``Sparse methods for
  direction-of-arrival estimation,'' in \emph{Academic Press Library in Signal
  Processing, Volume 7}.\hskip 1em plus 0.5em minus 0.4em\relax Elsevier, 2018,
  pp. 509--581.

\bibitem{khan2018angle}
A.~Khan, S.~Wang, and Z.~Zhu, ``Angle-of-arrival estimation using an adaptive
  machine learning framework,'' \emph{IEEE Communications Letters}, vol.~23,
  no.~2, pp. 294--297, 2018.

\bibitem{zhu2020two}
W.~Zhu, M.~Zhang, P.~Li, and C.~Wu, ``Two-dimensional doa estimation via deep
  ensemble learning,'' \emph{IEEE Access}, 2020.

\bibitem{kase2018doa}
Y.~Kase, T.~Nishimura, T.~Ohgane, Y.~Ogawa, D.~Kitayama, and Y.~Kishiyama,
  ``Doa estimation of two targets with deep learning,'' in \emph{2018 15th
  Workshop on Positioning, Navigation and Communications (WPNC)}.\hskip 1em
  plus 0.5em minus 0.4em\relax IEEE, 2018, pp. 1--5.

\bibitem{ahmed2020deep}
A.~M. Ahmed, U.~S.~K. Thanthrige, A.~E. Gamal, and A.~Sezgin, ``Deep learning
  for direction of arrival estimation via emulation of large antenna arrays,''
  \emph{arXiv preprint arXiv:2007.13824}, 2020.

\bibitem{pastorino2005smart}
M.~Pastorino and A.~Randazzo, ``A smart antenna system for direction of arrival
  estimation based on a support vector regression,'' \emph{IEEE transactions on
  antennas and propagation}, vol.~53, no.~7, pp. 2161--2168, 2005.

\bibitem{bialer2019performance}
O.~Bialer, N.~Garnett, and T.~Tirer, ``Performance advantages of deep neural
  networks for angle of arrival estimation,'' in \emph{ICASSP 2019-2019 IEEE
  International Conference on Acoustics, Speech and Signal Processing
  (ICASSP)}.\hskip 1em plus 0.5em minus 0.4em\relax IEEE, 2019, pp. 3907--3911.

\bibitem{liu2018direction}
Z.-M. Liu, C.~Zhang, and S.~Y. Philip, ``Direction-of-arrival estimation based
  on deep neural networks with robustness to array imperfections,'' \emph{IEEE
  Transactions on Antennas and Propagation}, vol.~66, no.~12, pp. 7315--7327,
  2018.

\bibitem{papageorgiou2021deep}
G.~K. Papageorgiou, M.~Sellathurai, and Y.~C. Eldar, ``Deep networks for
  direction-of-arrival estimation in low snr,'' \emph{IEEE Transactions on
  Signal Processing}, vol.~69, pp. 3714--3729, 2021.

\bibitem{tsoumakas2010random}
G.~Tsoumakas, I.~Katakis, and I.~Vlahavas, ``Random k-labelsets for multilabel
  classification,'' \emph{IEEE Transactions on Knowledge and Data Engineering},
  vol.~23, no.~7, pp. 1079--1089, 2010.

\bibitem{zhang2013review}
M.-L. Zhang and Z.-H. Zhou, ``A review on multi-label learning algorithms,''
  \emph{IEEE transactions on knowledge and data engineering}, vol.~26, no.~8,
  pp. 1819--1837, 2013.

\bibitem{kanters2020direction}
N.~B. Kanters, ``Direction-of-arrival estimation of an unknown number of
  signals using a machine learning framework,'' Master's thesis, University of
  Twente, 2020.

\bibitem{zheng2018feature}
A.~Zheng and A.~Casari, \emph{Feature engineering for machine learning:
  principles and techniques for data scientists}.\hskip 1em plus 0.5em minus
  0.4em\relax " O'Reilly Media, Inc.", 2018.

\bibitem{tsoumakas2007random}
G.~Tsoumakas and I.~Vlahavas, ``Random k-labelsets: An ensemble method for
  multilabel classification,'' in \emph{European conference on machine
  learning}.\hskip 1em plus 0.5em minus 0.4em\relax Springer, 2007, pp.
  406--417.

\bibitem{francois2017deep}
F.~Chollet, ``Deep learning with python,'' 2017.

\bibitem{goodfellow2016deep}
I.~Goodfellow, Y.~Bengio, A.~Courville, and Y.~Bengio, \emph{Deep
  learning}.\hskip 1em plus 0.5em minus 0.4em\relax MIT press Cambridge, 2016,
  vol.~1, no.~2.

\bibitem{kingma2014adam}
D.~P. Kingma and J.~Ba, ``Adam: A method for stochastic optimization,''
  \emph{arXiv preprint arXiv:1412.6980}, 2014.

\bibitem{abadi2016tensorflow}
M.~Abadi, A.~Agarwal, P.~Barham, E.~Brevdo, Z.~Chen, C.~Citro, G.~S. Corrado,
  A.~Davis, J.~Dean, M.~Devin \emph{et~al.}, ``Tensorflow: Large-scale machine
  learning on heterogeneous distributed systems,'' \emph{arXiv preprint
  arXiv:1603.04467}, 2016.

\bibitem{galar2011review}
M.~Galar, A.~Fernandez, E.~Barrenechea, H.~Bustince, and F.~Herrera, ``A review
  on ensembles for the class imbalance problem: bagging-, boosting-, and
  hybrid-based approaches,'' \emph{IEEE Transactions on Systems, Man, and
  Cybernetics, Part C (Applications and Reviews)}, vol.~42, no.~4, pp.
  463--484, 2011.

\end{thebibliography}

\begin{IEEEbiographynophoto}{Noud Kanters} 
\end{IEEEbiographynophoto}

\begin{IEEEbiographynophoto}{Andr\'{e}s Alay\'{o}n Glazunov} 
\end{IEEEbiographynophoto}

\end{document}